\documentclass[pra,twocolumn,superscriptaddress,showpacs,nobibnotes,notitlepage,nofootinbib]{revtex4-2}

\usepackage[utf8]{inputenc}
\usepackage[margin=1in]{geometry} 
\usepackage{amsmath,amsthm,amssymb,hyperref}
\hypersetup{
    colorlinks=true,
    linkcolor=blue,
    citecolor=magenta,
    filecolor=magenta,      
    urlcolor=blue,
    pdftitle={},
    pdfpagemode=FullScreen,
    }
\usepackage{adjustbox}
\usepackage{graphicx}
\usepackage{float}
\usepackage{xcolor}
\usepackage{soul}
\usepackage{qcircuit}

\usepackage{comment}
\usepackage{booktabs}

\definecolor{mycolor}{RGB}{0, 100, 200}

\begin{document}
\title{Real-time adaptation of quantum noise channel estimates}
\author{Lucas Daguerre}
\email{ldaguerre@ucdavis.edu}
\affiliation{Department of Physics and Astronomy, University of California, Davis, CA, 95616, USA}
\affiliation{Quantum Algorithms and Applications Collaboratory, Sandia National Laboratories, Livermore, CA 94550 USA}
\author{Mohan Sarovar}
\email{mnsarov@sandia.gov}
\affiliation{Quantum Algorithms and Applications Collaboratory, Sandia National Laboratories, Livermore, CA 94550 USA}
\date{\today}

\begin{abstract}
Estimates of noise channels for quantum gates are required for most error mitigation techniques and are desirable for informing quantum error correction decoders. These estimates can be obtained by resource-intensive off-line characterization techniques, but can become stale due to device drift and fluctuations. We propose a method to address this issue by performing real-time adaptation of noise channel estimates during the execution of a quantum algorithmic circuit using extended flag gadgets, mid-circuit measurements and Bayesian inference. We carry out analytical calculations and numerical simulations employing a Dirichlet prior distribution for the error rates in a Pauli channel to demonstrate and evaluate the technique, which can be seen as a protocol for real-time calibration of high-level gate error information.
\end{abstract}

\maketitle

\tableofcontents


\section{Introduction}
Precise understanding of noise in quantum computers is essential for the continued maturation of quantum computing technology and for extracting algorithmic utility from these devices. In the near-term, error mitigation techniques, such as probabilistic error cancellation (PEC) and Richardson extrapolation \cite{Temme2017-sc}, which aim to extend the computational utility of noisy quantum computers \cite{Kandala2019-zj}, benefit from precise characterization of device noise. Looking into the future, knowledge of device noise and its variation in time and space will be critical for efficient error decoding and correction in error-corrected and fault-tolerant devices. 

Although precise error characterization at the gate-level is possible using tools such as gate set tomography (GST) \cite{Nielsen2021-ex} and Pauli channel estimation \cite{Flammia2020-ce}, these techniques are time-consuming and are typically run infrequently. As a result, the descriptions of noise that they deliver might become inaccurate due to device and control drift during the execution of an algorithm. In fact, it has recently been noted that maintaining the accuracy of characterized noise models is challenging in the presence of device fluctuations such as those caused by uncontrolled two-level systems in superconducting devices, and that this limits the effectiveness of PEC \cite{Kim2024-af,Dasgupta2023-jk}. 

In this work, we propose a method to adapt noise channel estimates in \emph{real-time}, during the execution of an algorithm. The technique relies on intermediate measurements using \emph{extended flag gadgets}~\cite{Debroy2020,gonzales2023}, which provide information about gate errors while being non-intrusive and not interrupting algorithm execution. The idea of adapting noise channel estimates for more accurate PEC was also pursued by Dasgupta and Humble \cite{Dasgupta2023-jk}, however their technique relied on inference of noise parameter changes from a measurement of all qubits at the output of a circuit and so is not designed to non-intrusively adapt the noise channel estimates during execution of an algorithm. In addition, Gao \emph{et al.} have proposed the idea of using information from extended flag gadgets to collect coarse-grained error information about a quantum processor in real time \cite{Gao_2024}. While we discuss our real-time adaptation technique in the context of near-term devices and PEC, we note that the same principles could be used to track changes in device noise in early fault-tolerant, error-corrected devices and inform decoding algorithms. Finally, we note that our method could be viewed as a real-time calibration tool, where the calibration is not of device or control parameters, but of the description of the quantum channel induced by gate noise.

The rest of this article is organized as follows. Section~\ref{sec:background_material} provides some background material: Section~\ref{sec:error_model} describes the error model, and Section~\ref{sec:flagGadgets} introduces flag gadgets. Section~\ref{sec:Bayesian_inference} presents our main findings regarding noise channel adaptation using flag gadgets. This is further subdivided: Section~\ref{sec:Bayes_rule}, formulates Bayesian inference rules based on flag gadget measurements. Section~\ref{sec:Dirchlet}, describes the Dirichlet distribution used as the prior for the Bayesian inference. Section~\ref{sec:maxiam_case}, analyzes the special case when the number of flag gadget layers is maximal. Section~\ref{sec:genraliz_update}, provides exact update rules for generic flag gadgets, and Section~\ref{sec:approximae_rule} discusses an approximation to such exact inference rules. Section~\ref{sec:noisyGadgets}, studies the realistic scenario when noise is introduced by the flag gadgets.
Finally, Section~\ref{sec:conclusions} is a discussion of the results, possible extensions, and conclusion. Appendix~\ref{app:app_CNOT_all} and~\ref{app:alpha0} provide additional information about flag gadgets and the approximate update rule. Appendix~\ref{app:num_methods} provides information about the numerical methods.

\section{Background material}
\label{sec:background_material}
\subsection{Error model}
\label{sec:error_model}
To account for the presence of errors in a given quantum circuit, we model the implementation of every noisy gate acting on a quantum state $\rho$ as an ideal gate $U$ with action $\rho_U=U\rho U^{\dagger}$, followed by an error channel $\Lambda$,
\begin{equation}
    \Lambda(U\rho U^{\dagger})=\Lambda(\rho_U)\:.
    \label{eq:noisy_gate}
\end{equation}
The error channel $\Lambda$ can be generically represented as a completely-positive trace-preserving (CPTP) map, acting on $\rho_U$ as \cite{Nielsen_Chuang}
\begin{equation}
    \Lambda(\rho_U)=\sum_i E_i \rho_U E_i^{\dagger},\:\:\:\:\:\:\:\sum_i E_i^{\dagger}E_i=1\:,
    \label{eq:CPTP_map}
\end{equation}
with Kraus operators $E_i$, labeled by a natural number $i\in [1,4^{n_q}]$ for a $n_q$-qubit channel. In practice, we do not need to deal with such a generic error model. Indeed, error mitigation techniques such as PEC and Richardson extrapolation are less effective in the presence of such general error models. Instead, one typically uses randomized compiling \cite{wallman2016} to convert the error channel into a purely stochastic Pauli channel, 
\begin{equation}
    \tilde{\Lambda}(\rho_U)=\sum_i \lambda_i P_i\rho_U P_i^{\dagger},\:\:\:\:\:\:\:\sum_i \lambda_i=1\:,
\label{eq:Pauli_channel}
\end{equation}
with real coefficients $0 \leq \lambda_i \leq 1$ (again labeled by a natural number $i\in [1,4^{n_q}]$) and Pauli matrices $P_i$ belonging to the $n_q$-qubit Pauli group $\mathcal{P}_{n_q}=\{I,X,Y,Z\}^{\otimes n_q}$, defined as the tensor product of single-qubit Pauli matrices $I,X,Y,Z$. Throughout this article, our main objective will be to update in real-time the estimated values of the Pauli coefficients $\lambda_i$ for every noisy gate.

The mechanics of the Pauli twirling operation that underlies randomized compiling consist of applying elements of the Pauli group and their respective adjoint, before and after the error channel in place \cite{Dur2005,Emerson2007,Geller2013}
\begin{equation}
    \frac{1}{|\mathcal{P}_{n_q}|}\sum_{P_i \in \mathcal{P}_{n_q}}P_i^{\dagger}\Lambda(P_i \rho_U P_i^{\dagger})P_i  = \tilde{\Lambda}(\rho_U)\:,
\label{eq:pauli_twirl}
\end{equation}
where $|\mathcal{P}_{n_q}|=4^{n_q}$ is the cardinality of the Pauli group. 
Since we cannot apply the error channel by itself, but only the full noisy gate, the actual operational twirling procedure applies a Pauli gate $P_i^{\dagger}$ after the gate, and an ideal gate conjugated Pauli before the gate, $\tilde{P_i}=U^{\dagger}P_iU$:
\begin{equation}
\begin{split}
    \frac{1}{|\mathcal{P}_{n_q}|}\sum_{P_i \in \mathcal{P}_{n_q}}P_i^{\dagger} \Lambda(U\tilde{P}_i \rho \tilde{P}_i^{\dagger}U^{\dagger})P_i 
&=\tilde{\Lambda}(U\rho U^{\dagger})\\
&= \tilde{\Lambda} (\rho_U) \:.
\label{eq:twriling_effective}
\end{split}
\end{equation}
Notice that $\tilde{P_i}$ is a Pauli gate in case that $U$ is a Clifford gate \cite{Gottesman98,Nielsen_Chuang}. In randomized compiling these extra Pauli gates are compiled into the circuit such that the overall circuit depth does not increase \cite{wallman2016}.

\subsection{Flag gadgets}
\label{sec:flagGadgets}
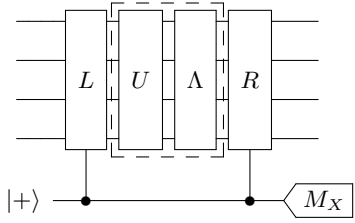
\begin{figure}[ht!]
\[
\begin{array}{c}
\Qcircuit @C=.5em @R=.7em  {              
&\qw &\qw &\qw          &\multigate{3}{L}&\multigate{3}{U}&\multigate{3}{\Lambda}&\multigate{3}{R}&\qw  \\
&\qw &\qw &\qw          &\ghost{L}       &\ghost{U}       &\ghost{\Lambda}       &\ghost{R}       &\qw       \\
&\qw &\qw &\qw          &\ghost{L}       &\ghost{U}       &\ghost{\Lambda}       &\ghost{R}       &\qw       \\
&\qw &\qw &\qw          &\ghost{L}       &\ghost{U}       &\ghost{\Lambda}       &\ghost{R}       &\qw        \\
&    &    &             &                &                &                      &                &   \\
&    &    &\lstick{|+\rangle}&\ctrl{-2}  &\qw             & \qw                  &\ctrl{-2}       &\measuretab{M_X}  \gategroup{1}{6}{4}{7}{.6em}{--}  \\
}
\end{array}
\]
\caption{Flag gadget acting on a noisy gate (dashed region: represented as a noiseless gate $U$ followed by a noise channel $\Lambda$). The gadget consist of an ancilla qubit prepared on the $|+\rangle$ state, followed by two controlled gates with unitaries left($L$)/right($R$), respectively, as well as a final measurement on the $X$-basis.} 
\label{fig:gadget_original} 
\end{figure}

The general purpose of this section is to revisit a method to extract information from the noise channel $\Lambda$  associated with a noisy gate, without disturbing the action of the gate. To accomplish this task, we utilize \textit{flag gadgets} \cite{Chao2018,Chamberland2018,Reichardt2021,Debroy2020,gonzales2023,Ewout2023,langfitt2024,Gao_2024}, which were originally introduced to flag/detect higher-weight errors introduced by stabilizer circuit measurements in quantum error correcting codes \cite{Chao2018,Chamberland2018,Reichardt2021}. The flag gadgets, see Fig.~\ref{fig:gadget_original}, consist of an auxiliary ancilla prepared in the state $|+\rangle$, along with two controlled gates labeled by the unitaries that act on the data qubits before ($L$) and after ($R$) an operation. Finally, the ancilla is measured in the $X$-basis. 

\begin{figure}[ht!]
\[
\begin{array}{c}
\Qcircuit @C=.5em @R=.7em  {              
&\qw &\multigate{3}{U}&\qw &\qw &\qw &\qw &\qw &\qw                            &\multigate{3}{R^{\dagger}}&\multigate{3}{\Lambda}&\multigate{3}{R}&\qw  \\
&\qw &\ghost{U}       &\qw &\qw &\qw &\qw &\qw &\qw                            &\ghost{R^{\dagger}}       &\ghost{\Lambda}       &\ghost{R}       &\qw       \\
&\qw &\ghost{U}       &\qw &\qw &\qw &\qw &\qw &\qw                            &\ghost{R^{\dagger}}       &\ghost{\Lambda}       &\ghost{R}       &\qw       \\
&\qw &\ghost{U}       &\qw &\qw &\qw &\qw &\qw &\qw                            &\ghost{R^{\dagger}}       &\ghost{\Lambda}       &\ghost{R}       &\qw       \\
&    &                &    &    &    &    &    &                               &                                    &                              &                          &       \\
&    &                &    &    &    &  |+\rangle   &   &             &\ctrl{-2}                           &\qw                           &\ctrl{-2}                 &\measuretab{M_X}  \gategroup{1}{6}{6}{13}{1.2em}{--}\\
  }
\end{array}
\]
\caption{Flag gadget equivalent to Fig.~\ref{fig:gadget_original}, due to the relation (\ref{eq:equivalence}). In case the error channel is also a Pauli channel $\tilde{\Lambda}$ (\ref{eq:Pauli_channel}), the dashed region acts as an effective Pauli channel $\tilde{\Lambda}_{\pm}(\rho_U;\vec{\lambda})$ (\ref{eq:Pauli_1layer}), where the dependence on the measurement outcome $m_X=\pm 1$ is emphasized.} \label{fig:gadget_equiv} 
\end{figure}
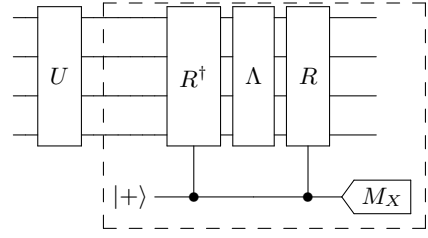

Given a noiseless unitary gate $U$, if the condition
\begin{equation}
    RUL=U\:,
\label{eq:equivalence}
\end{equation}
holds, the gate $U$ can be pushed through the left controlled gate in such a way that the gadget is effectively acting on the noise channel $\Lambda$, see Fig. \ref{fig:gadget_equiv}. 
For practical ease, we will choose $R$ to be a Pauli matrix, $R\in \mathcal{P}_{n_q}$. 
It is clear from Fig. \ref{fig:gadget_equiv} that if the noise is trivial, i.e., $\Lambda=I$, the action of the flag gadget is also trivial in the state $\rho$. Therefore, in a generic noisy scenario, we expect to answer a fundamental question: \textit{what information do the flag gadget measurements provide about the error channels?}  Initially, we will consider the ancilla preparation/measurement as well as the action of the controlled unitaries to be noiseless. We will discuss the impact of noisy flag gadgets in Section~\ref{sec:noisyGadgets}.

Upon introducing the extra ancilla in the plus state, the whole system is represented by the quantum state $\rho \otimes |+\rangle \langle +|$. Afterwards, the action of the flag gadget in Fig.~\ref{fig:gadget_original} on the initial state $\rho$, yields a post-measurement state $ \rho_{\text{post}}^{\pm}$ dependent on the measurement outcome $m_X=\pm 1$. In particular, if the error channel $\Lambda$ is described by a CPTP map (\ref{eq:CPTP_map}), the post-measurement state is \cite{gonzales2023}
\begin{equation}
\rho_{\text{post}}^{\pm}=\frac{\sum_i E_i^{\pm} \rho_U E_{i}^{\pm \dagger}}{\text{Tr}(\sum_i E_i^{\pm} \rho_U E_{i}^{\pm \dagger})},
    \label{eq:rho_post}
\end{equation}
where $\rho_U=U\rho U^{\dagger}$ and $E_i^{\pm}$ is defined as
\begin{equation}
    E_i^{\pm}=\frac{E_i  \pm  R E_i R^{\dagger}}{2}\:.
    \label{eq:E_pm}
\end{equation}


\begin{figure}[ht!]
    \[ \quad \:\:\:\:
    \begin{array}{c}
        \Qcircuit @C=0.32em @R=.4em {
            \lstick{}&\multigate{3}{L_3}&\multigate{3}{L_2}&\multigate{3}{L_1}&\multigate{3}{U}&\multigate{3}{\Lambda}&\multigate{3}{R_1}&\multigate{3}{R_2}&\multigate{3}{R_3}&\qw&\\
            \lstick{}&\ghost{L_3}       &\ghost{L_2}       &\ghost{L_1}       &\ghost{U}       &\ghost{\Lambda}       &\ghost{R_1}       &\ghost{R_2}       &\ghost{R_3}       &\qw&\\
            \lstick{}&\nghost{L_3}      &\nghost{L_2}      &\nghost{L_1}      &\nghost{U}      &\nghost{\Lambda}      &\nghost{R_1}      &\nghost{R_2}      &\nghost{R_3}      &\cdots &\\
            \lstick{}&\ghost{L_3}       &\ghost{L_2}       &\ghost{L_1}       &\ghost{U}       &\ghost{\Lambda}       &\ghost{R_1}       &\ghost{R_1}       &\ghost{R_2}       &\qw&\\
  \lstick{|+\rangle} &\qw               &\qw               &\ctrl{-1}         &\qw             &\qw                   &\ctrl{-1}         &\qw               &\qw               &\measuretab{M_X}  \\
  \lstick{|+\rangle} &\qw               &\ctrl{-2}         &\qw               &\qw             &\qw                   &\qw               &\ctrl{-2}         &\qw               &\measuretab{M_X}  \\
  \lstick{|+\rangle} &\ctrl{-3}         &\qw               &\qw               &\qw             &\qw                   &\qw               &\qw               &\ctrl{-3}         &\measuretab{M_X} \inputgroupv{1}{4}{1em}{2em}{n_q} \gategroup{1}{5}{4}{6}{.3em}{--}\\
        }
    \end{array}
    \]
    \caption{Example of a multi-layered flag gadget acting on a noisy $n_q$-qubit unitary (highlighted in the dashed region). In this case, there are $l=3$ layers. Again, we require $R_i U L_i = U$.}    \label{fig:gadgetGeneric}
\end{figure}
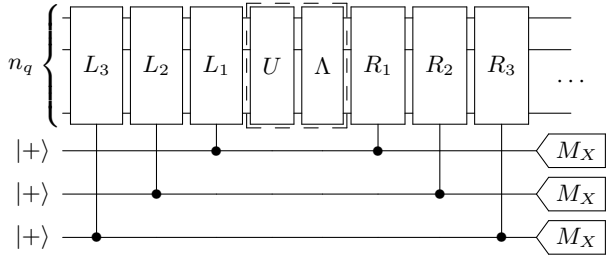

\noindent The probability of obtaining $\rho_{\text{post}}^{\pm}$ is, respectively,
\begin{equation}
    p(m_X=\pm 1)=\text{Tr}\left(\sum_i E_i^{\pm} \rho_U E_{i}^{\pm \dagger}\right)\:.
\label{eq:probability_rho}
\end{equation}
The effect of $E_i^{\pm}$ (\ref{eq:E_pm}) becomes more evident when expanding the Kraus operators $E_i$ in terms of Pauli matrices $P_j$
\begin{equation}
    E_i = \sum_j \Gamma_{ij}P_j\:,
\end{equation}
where $\Gamma_{ij}$ are complex coefficients. In case that the measurement is given by $m_X=+1$, the new operators are such that their Pauli components commute with $R$
\begin{equation}
    E_i^+=\sum_{j/[R,P_j]=0} \Gamma_{ij}P_j\:.
\end{equation}
Similarly, when $m_X=-1$ only those components that anti-commute with $R$ survive
\begin{equation}
    E_i^-=\sum_{j/\{R,P_j\}=0}\Gamma_{ij}P_j\:.
\end{equation}
We will focus on the case where the error channel $\Lambda$ is a Pauli channel (\ref{eq:Pauli_channel}), with $E_i=\sqrt{\lambda_i}P_i$.
In such a case, an interpretation of (\ref{eq:rho_post}) entails that the action of the flag gadget on the noisy gate (Fig.~\ref{fig:gadget_original}) is equivalent to the action of the noiseless gate $U$ followed by a modified Pauli channel $\tilde{\Lambda}_{\pm}$ (Fig.~\ref{fig:gadget_equiv}) with Kraus operators $E_i=\sqrt{\frac{\lambda_i}{\mathcal{N}}}P_i$,
\begin{equation}
    \tilde{\Lambda}_{\pm}(\rho_U;\vec{\lambda})=\frac{1}{\mathcal{N}} \sum_{ i\in [g]} \lambda_i P_i \rho_U P_i^{\dagger}\:,
\label{eq:Pauli_1layer}
\end{equation}
with a normalization factor given by (\ref{eq:probability_rho}),
\begin{equation}
    \mathcal{N}=\sum_{ i\in [g]} \lambda_i \:.
\label{eq:normalization}
\end{equation}
The set $[g]$ is defined such that it contains all the integers from the set $i\in [1,4^{n_q}]$ satisfying
\begin{equation}
    [g]=\big\{i:\:[P_i,R]_{b}=0\:,\: b=m_X\big\}\:,
\label{eq:set_g}
\end{equation}
where we have introduced the convention for the commutator ($b=+1$) and anti-commutator ($b=-1$), 
\begin{equation}
    [X,Y]_{b}=XY-b YX\:.
\label{eq:com_ant}
\end{equation}
This result is a consequence of the cyclic property of the trace $\text{Tr}(AB)=\text{Tr}(BA)$, that $E_i^{\pm \dagger}E_i^{\pm}=\lambda_i  I$ for a Pauli channel and that $\text{Tr}(\rho_U)=1$ since $U$ is unitary. 
Notably, the flag gadget construction of Fig.~\ref{fig:gadget_original} can be further generalized by adding multiple layers acting on the same noisy unitary \cite{gonzales2023}. In Fig.~\ref{fig:gadgetGeneric} we display an example of a $3$-layered flag gadget acting on a $n_q$-qubit noisy unitary.

Henceforth, we focus on the particular case where the error channel $\Lambda$ is a Pauli channel and $l$-layered gadgets are utilized. The results (\ref{eq:Pauli_1layer}) and (\ref{eq:normalization}) remain valid as long as the set $[g]$ (\ref{eq:set_g}) is redefined as the intersection of sets
\begin{equation}
    [g]=\bigcap_{q=1}^l \big\{i:\:[P_i,R_q]_{b_q}=0\:,\: b_q=m_X^{(q)}\big\}\:.
\label{eq:set_g_l}
\end{equation}
Thus, $P_i$ either must commute or anti-commute with the respective Pauli $R_q$ (depending of the measurement outcome of each gadget) to appear in the sum of (\ref{eq:Pauli_1layer}) and (\ref{eq:normalization}). In order to unclutter the notation, we decided to omit the $l$ dependence in the set notation $[g]$.

Note that for a $l$-layered gadget acting on a $n_q$-qubit noisy unitary, the maximum number of non-zero components of (\ref{eq:Pauli_1layer}) and (\ref{eq:normalization}) is $2^{2n_q-l}$. This is because originally there are $4^{n_q}=2^{2n_q}$ Pauli terms, and every layer $q=1,\dots,l$, removes half of them at a time. Finally, this result also implies that the maximal number of layers is given by $l_{\text{max}}=2n_q$, in which case the set $[g]$ contains a single element.

\section{Adaptation of noise channel estimates}
\label{sec:Bayesian_inference}
\subsection{Bayesian inference}
\label{sec:Bayes_rule}

In the previous section we saw how flag gadget measurements post-select out portions of a Pauli error channel inconsistent with the measurement results. Now we describe how this information can be used to update in real-time the estimates of the noise channel for a gate. We employ a Bayesian approach and accordingly treat the available characterization of the noise channel associated to a gate (available through off-line characterization) as a prior distribution over Pauli error rates. Namely, the preliminary knowledge of a Pauli channel $\tilde{\Lambda}$ associated to a noisy unitary gate can be encoded in a prior probability distribution $P(\vec{\lambda};\vec{\alpha})$, dependent on some variables $\vec{\lambda}=(\lambda_1,\cdots,\lambda_K)$ and hyperparameters $\vec{\alpha}=(\alpha_1,\cdots,\alpha_{K'})$. The expected values of the Pauli error rates under this prior distribution, $\{\langle \lambda_i\rangle_{P(\vec{\lambda};\vec{\alpha})}\}$, specify the expected value $\mathbb{E}\{\cdot\}$ of the Pauli channel $\tilde{\Lambda}$ as
\begin{equation}
\begin{split}
    \mathbb{E}\{\tilde{\Lambda}(\rho_U;\vec{\lambda})\} &=\int d\lambda P(\vec{\lambda};\vec{\alpha})\tilde{\Lambda}(\rho_U;\vec{\lambda})\\
    &=\sum_{i} \langle\lambda_i \rangle_{P(\vec{\lambda};\vec{\alpha})} P_i \rho_U P_i^{\dagger}\:.
\end{split}
\end{equation}
This description naturally allows the incorporation of characterization uncertainties in the form of a non-singular prior distribution over the Pauli error rates.
\textit{Bayes' rule} provides a procedure to incorporate the extra information acquired from the flag gadget measurement outcomes $m_X$ and update our description of the error channel. Namely, Bayes' rule gives a formula for the posterior probability $P(\vec{\lambda}|m_X)$ after incorporation of the measurement outcomes $m_X$, 
\begin{equation}
\displaystyle
P( \vec{\lambda}|m_X)=\frac{P(m_X| \vec{\lambda})P( \vec{\lambda};\vec{\alpha})}{P(m_X)}\:.
\label{eq:Bayes_eqn}
\end{equation}
Here, the likelihood $P(m_X | \vec{\lambda})$ gives the conditional probability of obtaining a measurement outcome $m_X$ from the given prior distribution. The marginal probability $P(m_X)$, gives the unconditional probability of measuring $m_X$.

\begin{figure}[ht!]
\[
\quad \:\:
\begin{array}{c}
\Qcircuit @C=0.2em @R=.5em   {                 
&\qw &\multigate{1}{\tilde{L}_1}&\multigate{1}{V}&\multigate{1}{\tilde{R}_1}&\cds{1}{\cdots}&\cds{1}{\cdots}  &\qw               &\multigate{1}{\tilde{L}_n}&\multigate{1}{V}&\multigate{1}{\tilde{R}_n}&\qw \\
 &\qw &\ghost{\tilde{L}_1}       &\ghost{V}      &\ghost{\tilde{R}_1}       &\qw            &\qw              &\qw               &\ghost{\tilde{L}_n}        &\ghost{V}      &\ghost{\tilde{R}_n}       &\qw           \\
\lstick{|+\rangle}&\qw &\ctrl{-1}         &\qw              &\ctrl{-1}         &\measuretab{M_X}&                &\lstick{|+\rangle}&\ctrl{-1}        &\qw           &\ctrl{-1}         &\measuretab{M_X}   }          
\end{array}
\]
\caption{Repeated action of a noisy gate $V$  $n$-consecutive times. For each application of the noisy gate $V$, a set of $l$-layered flag gadgets is measured (the tildes indicate that more than a single layer could be measured per step).} \label{fig:gadgets_repeated} 
\end{figure}
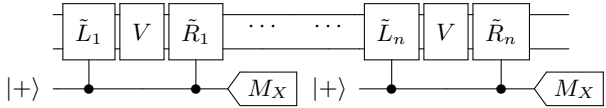

The likelihood is the quantum-mechanical probability of obtaining the post-measurement state $\rho_{\text{post}}^{\pm}$ (\ref{eq:probability_rho}), respectively,
\begin{equation}
\begin{split}
   P(m_X| \vec{\lambda})=p(m_X)=\sum_{ i\in [g]} \lambda_i\:,
\label{eq:likelihood_orig}
\end{split}
\end{equation}
which is by definition equal to the normalization factor of (\ref{eq:normalization}).
The marginal probability is then,
\begin{equation}
\begin{split}
    P(m_X)&= \int d \lambda\: P(m_X| \vec{\lambda}) P(\vec{\lambda};\vec{\alpha})\\
    &=\sum_{i\in [g]}\langle \lambda_i \rangle_{P(\vec{\lambda};\vec{\alpha})}\:.
\end{split}
\end{equation}
Thus, the Bayes update rule is
\begin{equation}
    P( \vec{\lambda}|m_X)=\left(\frac{\displaystyle\sum_{i\in [g]}\lambda_i}{\displaystyle\sum_{i\in [g]}\langle \lambda_i \rangle_{P(\vec{\lambda};\vec{\alpha})}}\right) P(\vec{\lambda};\vec{\alpha})\:.
\label{eq:Bayes_equation}
\end{equation}
The equation (\ref{eq:Bayes_equation}) provides an update rule for the expected values upon measuring $m_X$,
\begin{equation}
    \langle \lambda_j\rangle_{P(\vec{\lambda};\vec{\alpha})} \longrightarrow \langle \lambda_j\rangle_{P( \vec{\lambda}|m_X)} \:.
\label{eq:expected_general}
\end{equation}
A useful scenario for the application of the update rule (\ref{eq:expected_general}) requires the repetition of the flag gadget measurements. Hence, we focus on the particular case where a given noisy unitary $V$ acts repeatedly on the same subset of qubits. In addition, we also assume that the error channel $\tilde{\Lambda}$ does not drift with time (assumption that is valid if the time it takes to apply the full sequence of flag gadgets is much smaller that the characteristic drift time). Under these assumptions, we can apply a sequence of $l$-layered gadgets $n$ consecutive times as in Fig~\ref{fig:gadgets_repeated}. 

We treat the posterior of the $k$-th step as the prior of the $(k+1)$-th step, for $k=1,\dots,n$. Hence, we can write the recursive relation
\begin{equation}
   P^{(k)}(\vec{\lambda}) =\frac{\big(\sum_{i_k\in [g_k]}  \lambda_{i_k} \big) }{\big(\sum_{i_k\in [g_k]} \langle \lambda_{i_k}\rangle_{P^{(k-1)}} \big) }P^{(k-1)}(\vec{\lambda})\:.
\label{eq:recursive_bayes}
\end{equation}
Here $\langle \cdot \rangle_{P^{(k-1)}}$ indicates that the expected values are computed with respect to the prior distribution $P^{(k-1)}(\vec{\lambda})$, where we suppressed the dependence on the hyperparameters $\vec{\alpha}$ for notational simplicity. The initial distribution $P^{(0)}(\vec{\lambda})$ corresponds to the prior distribution $P(\vec{\lambda};\vec{\alpha})$. In addition, the set $[g_k]$ stands for the set $[g]$ (\ref{eq:set_g_l}) obtained by using a $l$-layered gadget at the $k$-th step (thus, we only emphasize the step number and not the number of layers with this notation). 

Using (\ref{eq:recursive_bayes}), we can compute the expected values after $k$-steps as
\begin{equation}
\langle \lambda_j \rangle_{P^{(k)}}=\frac{\big(\sum_{i_k\in [g_k]}  \langle \lambda_j\lambda_{i_k}\rangle_{P^{(k-1)}} \big) }{\big(\sum_{i_k\in [g_k]} \langle \lambda_{i_k}\rangle_{P^{(k-1)}} \big) }\:.
\label{eq:expected_raw}
\end{equation}
The numerator of (\ref{eq:expected_raw}) can be rewritten by recalling the definition of covariance,
\begin{equation}
\begin{split}
    \langle \lambda_j\lambda_{i_k}\rangle_{P^{(k-1)}}&=\langle \lambda_j\rangle_{P^{(k-1)}}\langle \lambda_{i_k}\rangle_{P^{(k-1)}}\\
    &\quad\quad +\text{Cov}(\lambda_j,\lambda_{i_k})_{P^{(k-1)}}\:.
\end{split}
\end{equation}
Therefore the recursive update rule can be recast as
\begin{equation}
\begin{split}
    \langle \lambda_j \rangle_{P^{(k)}}&=\langle \lambda_j \rangle_{P^{(k-1)}}\\
    &\quad \quad +\frac{\big(\sum_{i_k\in [g_k]}  \text{Cov}(\lambda_j,\lambda_{i_k})_{P^{(k-1)}} \big) }{\big(\sum_{i_k\in [g_k]} \langle \lambda_{i_k}\rangle_{P^{(k-1)}} \big) }\:.
\label{eq:recursive_covariance}
\end{split}
\end{equation}
Intuitively (\ref{eq:recursive_covariance}) shows that if $\lambda_j$ is correlated with $\lambda_{i_k}$ the expected value increases, whereas if it is anti-correlated, the expected value decreases. For example, in the next subsection we specialize to the Dirichlet distribution as the prior, where we compute the covariance and means explicitly (\emph{e.g.,} (\ref{eq:cov_dirich}) and (\ref{eq:lmbda_1})), and note that $\lambda_j$ is correlated with $\lambda_{i_k}$ when $j \in [g_k]$ and anti-correlated when $j \notin [g_k]$. 

Finally, equation (\ref{eq:recursive_bayes}) can also be applied $n$-consecutive times starting from the original prior distribution $P^{(0)}(\vec{\lambda})\equiv P(\vec{\lambda};\vec{\alpha})$. This gives the probability distribution at the $n$-the step in terms of the probability distribution at the zero-th step
\begin{equation}
   P^{(n)}(\vec{\lambda}) =\frac{\prod_{k=1}^n \big(\sum_{i_k\in [g_k]}  \lambda_{i_k} \big)}{\prod_{k=1}^n \big(\sum_{i_k\in [g_k]} \langle \lambda_{i_k}\rangle_{P^{(k-1)}} \big)}P^{(0)}(\vec{\lambda})\:.
\label{eq:order_n_equation}
\end{equation}
Furthermore, using (\ref{eq:order_n_equation}) we can write down a generic update rule in terms of higher moments of the original prior distribution
\begin{equation}
    \langle \lambda_j \rangle_{P^{(n)}}=\frac{\sum_{i_n\in [g_n]} \ldots \sum_{i_1\in [g_1]}  \langle \lambda_{j}\lambda_{i_n}\ldots \lambda_{i_1}\rangle_{P^{(0)}}}{\sum_{i_n\in [g_n]} \ldots \sum_{i_1\in [g_1]}  \langle \lambda_{i_n}\ldots \lambda_{i_1}\rangle_{P^{(0)}}}\:.
\label{eq:update_general_expected}
\end{equation}
We explore this route further in Section~\ref{sec:genraliz_update}.
\subsection{Dirichlet distribution}
\label{sec:Dirchlet}

The treatment so far has been for arbitrary prior distributions over the Pauli error rates, $\lambda_i$. Now we specialize to a particularly useful prior distribution. 
First, we note that a sensible prior distribution must accommodate certain characteristics. It has to be a multi-variable continuous probability distribution over bounded $\lambda_i \in [0,1]$, and ideally should also incorporate the normalization constraint $\sum_i \lambda_i=1$.
A natural candidate that satisfies all these features is the Dirichlet distribution \cite{Balakrishnan2003}.

\begin{figure}[ht]
    \centering
    \includegraphics[width=0.48\textwidth]{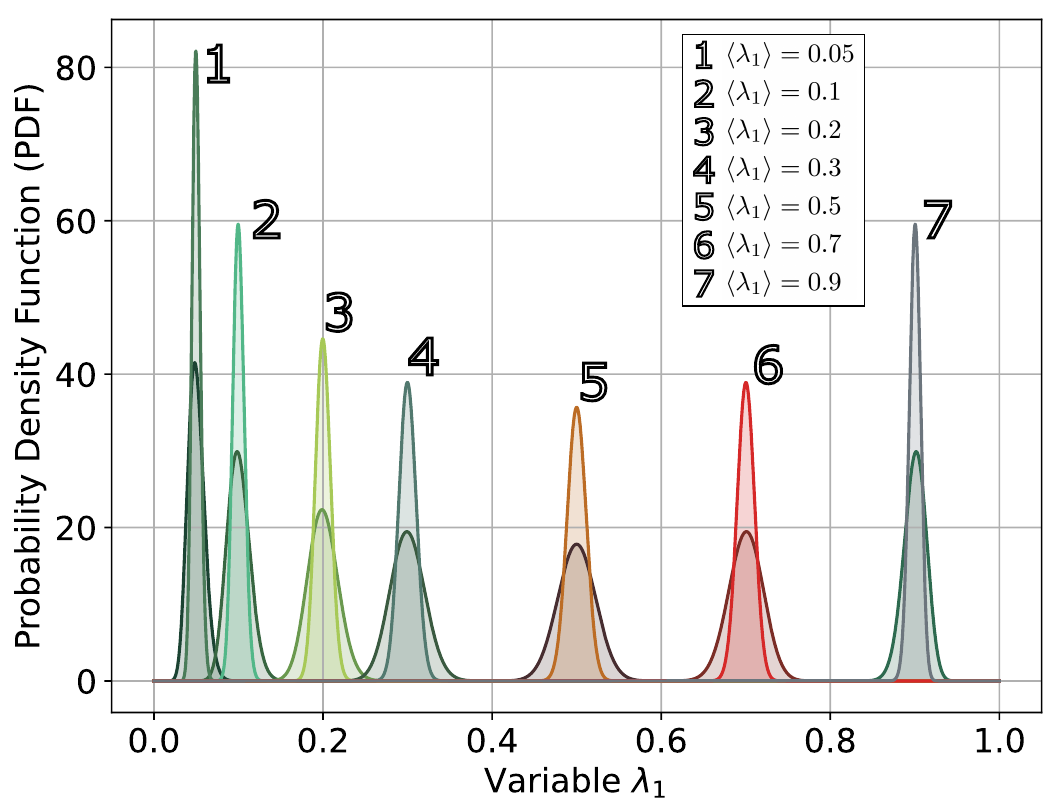}
    \caption{Numerical examples of the probability density function (PDF) for $14$ different Dirichlet distributions with $K=2$ variables (Beta distributions). For each of the different seven peaks the coefficient $(\alpha_1,\alpha_2)$ were obtained from different values of $\langle \lambda_1\rangle$ and $\alpha_0$. In particular, for a given peak number, the taller PDF corresponds to $\alpha_0=2000$ and the shorter to $\alpha_0=500$. Consistent with (\ref{eq:var_Dir}), larger values of $\alpha_0$ give rise to a narrower distribution.}
    \label{fig:beta_distr}
\end{figure}

The Dirichlet distribution depends on $K\geq 2$ variables $\vec{\lambda}=(\lambda_1,\dots,\lambda_{K})$ and positive real parameters $\vec{\alpha}=(\alpha_1,\cdots,\alpha_{K})$. It is explicitly defined as
\begin{equation}
D(\vec{\lambda};\vec{\alpha})=\frac{1}{B(\vec{\alpha})}\prod_{i=1}^{K}\lambda_i^{\alpha_i-1}\:\:\:\:\:\:,\:\:\:\:\:\: \sum_{i=1}^{K}\lambda_i=1\:.
\label{eq:Dir_distr}
\end{equation}
Note that that there are only $K-1$ independent variables. The denominator is given by the generalized Beta function, which is defined as
\begin{equation}
B(\vec{\alpha})=\frac{\Gamma(\alpha_1)\cdots 
  \Gamma(\alpha_{K})}{\Gamma(\alpha_1+\dots+\alpha_{K})}\:,
\label{eq:gen_beta_function}
\end{equation}
where $\Gamma(\cdot)$ is the Gamma function. When $K=2$ the Dirichlet distribution becomes the Beta distribution. For a Pauli channel acting on $n_q$-qubits (\ref{eq:Pauli_channel}), $K=4^{n_q}$. In Fig.~\ref{fig:beta_distr}, we show some plots of the Dirichlet distribution for the special case $K=2$. The formula for the expected values is
\begin{equation}
    \langle \lambda_i \rangle_{D(\vec{\lambda};\vec{\alpha})}=\frac{\alpha_i}{\alpha_0},
\label{eq:exp_Dir}
\end{equation}
and for the variance
\begin{equation}
    \text{Var}(\lambda_j)_{D(\vec{\lambda};\vec{\alpha})}=\frac{\langle \lambda_j \rangle_{D(\vec{\lambda};\vec{\alpha})} (1-\langle \lambda_j \rangle_{D(\vec{\lambda};\vec{\alpha})})}{\alpha_0+1}\:.
\label{eq:var_Dir}
\end{equation}
Here the sum over $\alpha_j$ is denoted as 
\begin{equation}
    \alpha_0=\sum_{j=1}^{K}\alpha_j\:.
\end{equation}

The Dirichlet distribution can be completely determined by specifying the set of expected values $\{\langle \lambda_i \rangle_{D(\vec{\lambda};\vec{\alpha})}\}$ in conjunction with $\alpha_0$. This also implies that the variance of every $\lambda_i$ are non-trivially related since
\begin{equation}
    \frac{\text{Var}(\lambda_j)_{D(\vec{\lambda};\vec{\alpha})}}{\text{Var}(\lambda_i)_{D(\vec{\lambda};\vec{\alpha})}}=\frac{\langle \lambda_j \rangle_{D(\vec{\lambda};\vec{\alpha})} (1-\langle \lambda_j \rangle_{D(\vec{\lambda};\vec{\alpha})})}{\langle \lambda_i \rangle_{D(\vec{\lambda};\vec{\alpha})} (1-\langle \lambda_i \rangle_{D(\vec{\lambda};\vec{\alpha})})}\:.
\end{equation}
An interesting property of the Dirichlet distribution is that the covariance has a closed form expression
\begin{equation}
\begin{split}
    \text{Cov}(\lambda_i,\lambda_j)_{D(\vec{\lambda};\vec{\alpha})}&=\delta_{i,j}\frac{\langle \lambda_j\rangle_{D(\vec{\lambda};\vec{\alpha})}}{\alpha_0+1}\\
    &\quad  -\frac{\langle \lambda_i\rangle_{D(\vec{\lambda};\vec{\alpha})}\langle \lambda_j\rangle_{D(\vec{\lambda};\vec{\alpha})}}{\alpha_0+1}\:,
\label{eq:cov_dirich}
\end{split}
\end{equation}
where $\delta_{i,j}$ is the Kronecker delta. Hence (\ref{eq:cov_dirich}) allows us to extract an exact analytic update rule for $\langle \lambda_j\rangle_{P^{(1)}}$ using (\ref{eq:recursive_covariance}) when $k=1$ and the prior is a Dirichlet distribution  $P^{(0)}(\vec{\lambda})=D(\vec{\lambda};\vec{\alpha})$,
\begin{equation}
    \langle \lambda_j \rangle_{P^{(1)}}=\frac{\alpha_0 \langle \lambda_j \rangle_{P^{(0)}}}{\alpha_0+1}+\frac{\delta(j,[g_1])}{\alpha_0+1}\frac{\langle \lambda_j \rangle_{P^{(0)}}}{\sum_{i_1 \in [g_1]}\langle \lambda_{i_1}\rangle_{P^{(0)}}}\:.
\label{eq:lmbda_1}
\end{equation}
In (\ref{eq:lmbda_1}) we utilized the generalized delta function defined as
\begin{equation}
    \delta(j,[g_k])=\begin{cases}
        1\quad \text{if}\quad j\in [g_k]\\
        0\quad \text{if}\quad j\notin [g_k]\\
    \end{cases}\:.
\label{eq:generalized_delta}
\end{equation}
In the following two subsections we will demonstrate flag gadget-based adaptation of noise channels when the prior distribution over Pauli error rates is Dirichlet. 

\subsection{Update rule: maximal number of layers}
\label{sec:maxiam_case}
In this subsection we study the special case when the number of flag gadget layers is maximal $l=2n_q$. As suggested by \cite{gonzales2023}, the general $n_q$-qubit case entails utilizing for the right unitaries $R_1,\cdots ,R_{2n_q}$ all single qubit $X$ and $Z$ gates acting on every data qubit. The left unitaries are then determined by the formula (\ref{eq:equivalence}).  We show an example of a $2$-layered gadget for a single qubit unitary in Fig.~\ref{fig:single_qubit_gadget}, and of a $4$-layered gadget for a CNOT gate in Appendix~\ref{app:4layeredCNOT}.
\begin{figure}[ht!]
\[
\begin{array}{c}
\Qcircuit @C=1.2em @R=.5em  {                   &\gate{U_Z}  &\gate{U_X}   & \gate{U} &\gate{\tilde{\Lambda}} &\gate{X}   &\gate{Z}  &\qw               &\\
\lstick{|+\rangle} &\qw       & \ctrl{-1} &  \qw                   & \qw&\ctrl{-1} &\qw       & \measuretab{M_X} & m_X^{(1)}\\
\lstick{|+\rangle} &\ctrl{-2} & \qw       & \qw                    & \qw  &\qw     &\ctrl{-2} & \measuretab{M_X} & m_X^{(2)} \gategroup{1}{4}{1}{5}{.6em}{--}\\
}
\end{array}
\]
\caption{Single qubit noisy gate with a 2-layered gadget. The dashed region comprises of a noiseless gate $U$ followed by a noise channel $\tilde{\Lambda}$. The unitaries are $U_P=UPU^{\dagger}$ for $P=X,Z$, which are also Pauli matrices when $U$ is a Clifford gate.} \label{fig:single_qubit_gadget} 
\end{figure}
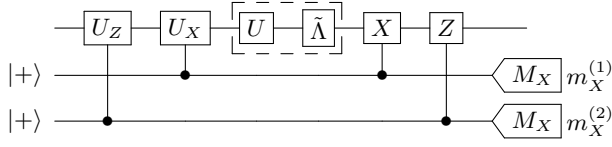

We use the Dirichlet distribution (\ref{eq:Dir_distr}) as the prior distribution
\begin{equation}
    P^{(0)}(\vec{\lambda})=D(\vec{\lambda};\vec{\alpha})\:.
\end{equation}
As observed earlier, using the maximal number of flag gadget layers, the post-selection and likelihood function singles out an individual $\lambda_j$ in (\ref{eq:recursive_bayes}), because each layer eliminates half of the Pauli rates in the effective error channel (\ref{eq:Pauli_1layer}). An example of this result can be seen in Table~\ref{tab:likelihood_singleq}.
\begin{table}[!ht]
\centering
\begin{tabular}{|c|c|c|}
\hline
$m_X^{(1)}$ & $m_X^{(2)}$ & $P(m_X|\vec{\lambda})$ \\ \hline
$+1$        & $+1$        & $\lambda_I$                               \\ \hline
$+1$        & $-1$        & $\lambda_X$                               \\ \hline
$-1$        & $-1$        & $\lambda_Y$                               \\ \hline
$-1$        & $+1$        & $\lambda_Z$                               \\ \hline
\end{tabular}
\caption{Likelihood probability of measuring $m_X$ given a prior probability distribution $P(\vec{\lambda})$ for a $2$-layered flag gadget acting on a single qubit unitary. The right column was constructed using (\ref{eq:set_g_l}), and indices $i\in [1,4]$ where replaced by $i \in \{I,X,Y,Z\}$, respectively.}
\label{tab:likelihood_singleq}
\end{table}

In case that $\lambda_j$ is singled out by the measurement outcome $m_X$ (i.e., $j \in [g]$ (\ref{eq:set_g_l})) Bayes' equation (\ref{eq:recursive_bayes}) can be written as
\begin{equation}
P^{(1)}(\vec{\lambda}) = \frac{\lambda_j}{\langle \lambda_j \rangle_{D(\vec{\lambda};\vec{\alpha})}}D(\vec{\lambda};\vec{\alpha}) = D(\vec{\lambda};\vec{\alpha}')\:.
\label{eq:bayes_single}
\end{equation}
As shown, $P^{(1)}(\vec{\lambda})=D(\vec{\lambda};\vec{\alpha}')$ is also a Dirichlet distribution for this particular case. Indeed, the exact update rule for the hyperparameters $\vec{\alpha}$ is
\begin{equation}
    \quad \alpha_j'=\alpha_j+1 \:\quad,\:\quad \alpha_i'=\alpha_i \:\:\quad\text{for}\:\:i\neq j\:,
\label{eq:bayes_maximal}
\end{equation}
or equivalently,
\begin{equation}
\alpha_i'=\alpha_i+\delta(i,[g])\:.
\label{eq:bayes_max_deltaf}
\end{equation}
This remarkable result follows from the recursive property of the Gamma function $\Gamma(x+1)=x\Gamma(x)$ and the expression for the expected values (\ref{eq:exp_Dir}). Indeed, if $\vec{e}_j$ represents a unit vector whose only non-zero component is in the $j$-th position, then
\begin{equation}
\begin{split}    B(\vec{\alpha}+\vec{e}_j)&=\frac{\Gamma(\alpha_1)\cdots \Gamma(\alpha_j+1) \cdots \Gamma(\alpha_{K})}{\Gamma(\alpha_1+\dots+(\alpha_j+1)+\dots+\alpha_{K})}\\
&=\frac{\alpha_j}{\alpha_0}\frac{\Gamma(\alpha_1)\cdots \Gamma(\alpha_j) \cdots \Gamma(\alpha_{K})}{\Gamma(\alpha_1+\dots+\alpha_j+\dots+\alpha_{K})}\\
&=\langle \lambda_j \rangle_{D(\vec{\lambda};\vec{\alpha})} B(\vec{\alpha})\:.
\label{eq:proof_propert_gamma}
\end{split}
\end{equation}
Given the exact analytic update rule for the Dirichlet distribution (\ref{eq:bayes_maximal}), we can also write down an exact analytic update rule for the expected values after $n$ consecutive steps
\begin{equation}
    \langle \lambda_j \rangle_{P^{(n)}}=\frac{\alpha_0 }{\alpha_0+n}\langle \lambda_j \rangle_{P^{(0)}}+\frac{1}{\alpha_0+n}\sum_{t=1}^{n}\delta(j,[g_t])\:,
\label{eq:expected_4gadgets}
\end{equation}
Recall that the generalized delta function was defined in (\ref{eq:generalized_delta}). Note that (\ref{eq:expected_4gadgets}) corresponds to applying $n$ consecutive times the equation (\ref{eq:lmbda_1}) when the cardinality $|[g_i]|=1$ for $i=1,\dots,n$. 

\begin{figure*}[ht]
    \centering
    \includegraphics[width=0.80\textwidth]{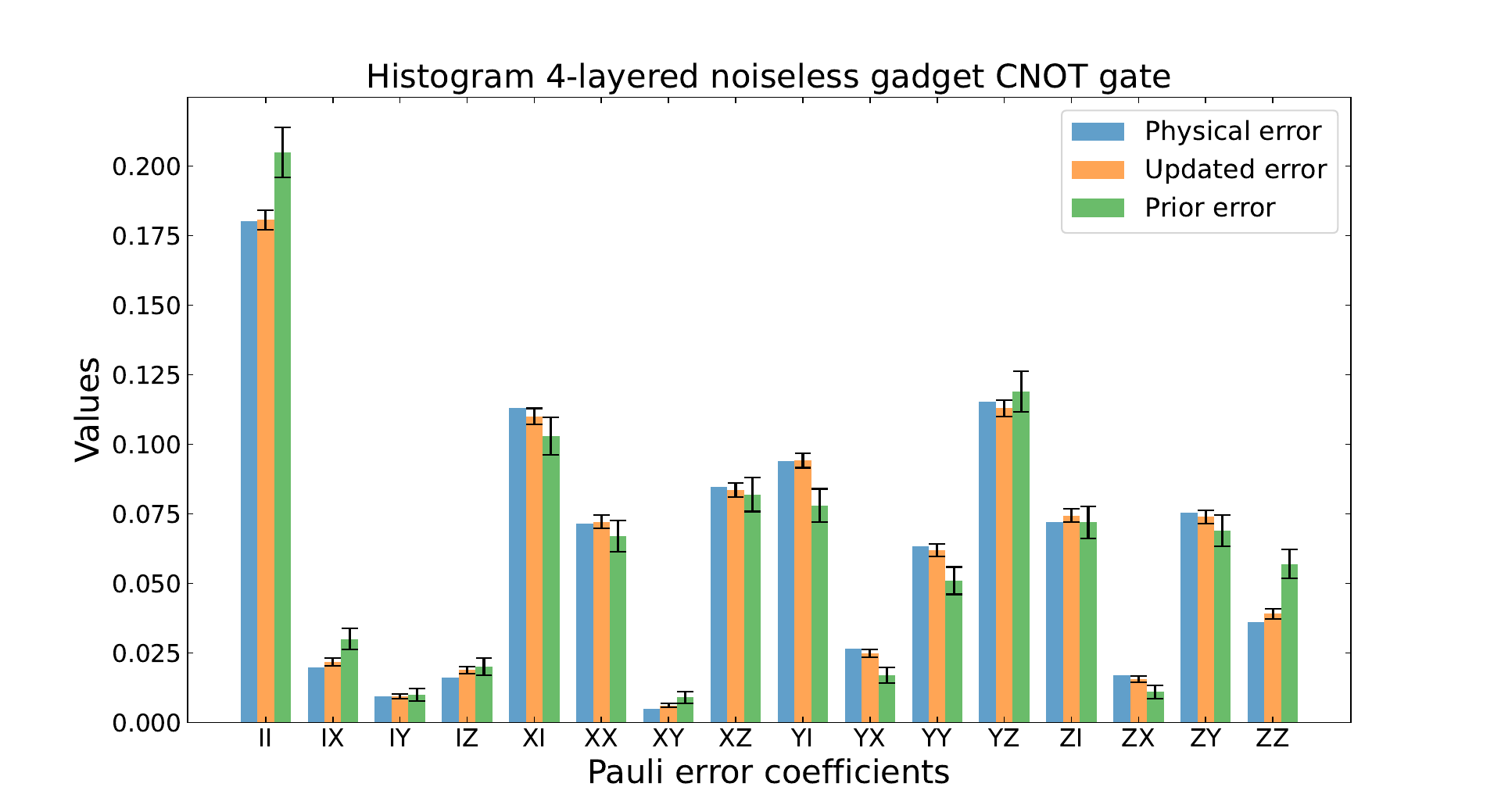}
    \caption{Histogram of the Pauli error rate estimates associated to a CNOT gate using noiseless $4$-layered gadgets. The `prior error' and the `updated error' columns display the expected values under the prior distribution $\langle \lambda_i\rangle_{P^{(0)}}$, and the updated distribution $\langle \lambda_i\rangle_{P^{(n)}}$ (\ref{eq:expected_4gadgets}), respectively. The `physical error' column represents the error rates $\langle \lambda_i\rangle_{P_{\text{phy}}}$ of the real Pauli channel $\tilde{\Lambda}$ associated to the CNOT gate. To simulate device drift, the `physical error' values slightly deviate from the `prior error' estimates, hence by utilizing the flag gadgets we aim to learn their differences in real-time. The numerical simulations were done considering the set up of Fig.~\ref{fig:gadgets_repeated} starting from the $|00\rangle$ state, taking $n=10000$ consecutive updates and $\alpha_0=2000$, and using the stabilizer circuit simulator package stim \cite{gidney2021stim}. The error bars represent one standard deviation obtained from the square root of (\ref{eq:var_4layered}). Details of the $4$-layered flag gadgets for a CNOT gate are described in Appendix~\ref{app:4layeredCNOT}.}
    \label{fig:4layers_CNOT}
\end{figure*}

Furthermore, because the updated distribution is also a Dirichlet distribution, the variance after $n$ steps is
\begin{equation}
    \text{Var}(\lambda_j)_{P^{(n)}}=\frac{\langle \lambda_j\rangle_{P^{(n)}}(1-\langle \lambda_j\rangle_{P^{(n)}})}{\alpha_0+n+1}\:.
\label{eq:var_4layered}
\end{equation}
A numerical simulation utilizing the update rule (\ref{eq:expected_4gadgets}) for a 4-layered gadget acting on a $\text{CNOT}$ gate is shown in Fig.~\ref{fig:4layers_CNOT}. As this figure shows, the error rates are updated to reflect the new error channel. 

\begin{figure*}[ht]
    \centering
    \includegraphics[width=0.80\textwidth]{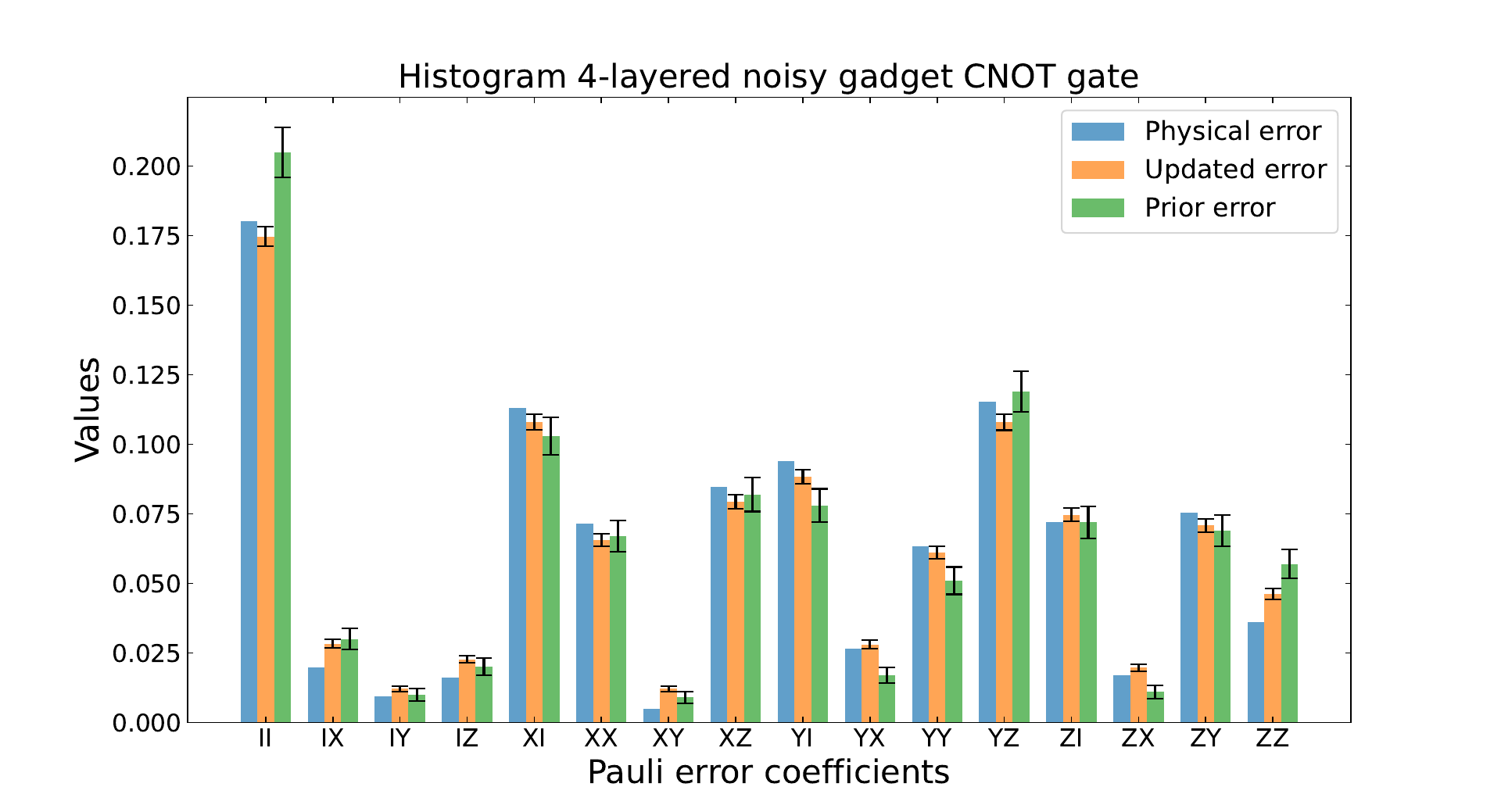}
    \caption{Histogram of the Pauli error estimates associated to a CNOT gate using noisy $4$-layered gadgets. The noise introduced by the flag gadgets was modeled using a uniform depolarizing error model with strength $p=0.01$ (described in Section~\ref{sec:noisyGadgets}). Otherwise, the simulations were done as in Fig.~\ref{fig:4layers_CNOT}.}
    \label{fig:4layers_CNOT_noisy}
\end{figure*}

\subsection{Generalized update rule}
\label{sec:genraliz_update}
In this subsection we study the generic update rule case in which the number of layers is not necessarily maximal. In terms of practical implementation, this is the most relevant case. Recall that at the end of the day, the gadgets do also introduce additional noise, hence reducing at maximum the total number of implemented gadgets is crucial.  

Following the lines of the generic expression for the expected values obtained after $n$ iterations of the $l$-layered gadgets (\ref{eq:update_general_expected}), it turns out that the expected values can be written in close form. This is because high-moments of the Dirichlet distribution are determined in terms of the generalized Beta function (\ref{eq:gen_beta_function}), with a formula given by \cite{Balakrishnan2003},
\begin{equation}
    \langle \lambda_{i_k}\ldots \lambda_{i_1}\rangle_{P^{(0)}}=\frac{B(\vec{\alpha}+\vec{e}_{i_1}+\dots+\vec{e}_{i_k})}{B(\vec{\alpha})}\:.
\label{eq:beta_expectec}
\end{equation}
Note that the result (\ref{eq:beta_expectec}) recovers the single variable case proven in (\ref{eq:proof_propert_gamma}),
\begin{equation}
     \langle \lambda_{j}\rangle_{P^{(0)}}=\frac{B(\vec{\alpha}+\vec{e}_j)}{B(\vec{\alpha})}\:.
\end{equation}
However, the number of terms to be summed in (\ref{eq:update_general_expected}) is $O(|[g_k]|^{n})$, where $|[g_k]|$ is the cardinality of the set $[g_k]$. This situation suggests that the computation is numerically intractable in its full form. To better understand this issue, let us analyze the two possible cases: i) the cardinality is $|[g_k]|=1$, which corresponds to the maximal number of gadgets studied in Section~\ref{sec:maxiam_case}. In this case, the distribution remains a Dirichlet distribution throughout every iteration; ii) the cardinality is $|[g_k]|>1$. In such an event, after $n\geq 1$ iterations, the distribution becomes a convex combination of Dirichlet distributions. To illustrate this point, suppose that the first measurement outcome $m_X$ selects $[g_1]=\{i,j\}$ with $i\neq j$, the posterior distribution is then
\begin{equation}
\begin{split}    P(\vec{\lambda}|m_X)&=\frac{\lambda_i+\lambda_j}{\langle \lambda_i\rangle_{D(\vec{\lambda};\vec{\alpha})}+\langle \lambda_j \rangle_{D(\vec{\lambda};\vec{\alpha})}}D(\vec{\lambda};\vec{\alpha})\\
&=\frac{\langle\lambda_i\rangle_{D(\vec{\lambda};\vec{\alpha})}}{\langle \lambda_i\rangle_{D(\vec{\lambda};\vec{\alpha})}+\langle \lambda_j \rangle_{D(\vec{\lambda};\vec{\alpha})}}D(\vec{\lambda};\vec{\alpha}+\vec{e}_i)\\
&\quad +\frac{\langle\lambda_j\rangle_{D(\vec{\lambda};\vec{\alpha})}}{\langle \lambda_i\rangle_{D(\vec{\lambda};\vec{\alpha})}+\langle \lambda_j \rangle_{D(\vec{\lambda};\vec{\alpha})}}D(\vec{\lambda};\vec{\alpha}+\vec{e}_j).
\end{split}
\end{equation}
Thus, generically the number of terms describing the probability distribution also grows exponentially in the number of iterations as $O(|[g_k]|^{n})$.
\subsubsection{Approximate update rule}
\label{sec:approximae_rule}
\begin{figure*}[ht]
    \centering
    \includegraphics[width=0.8\textwidth]{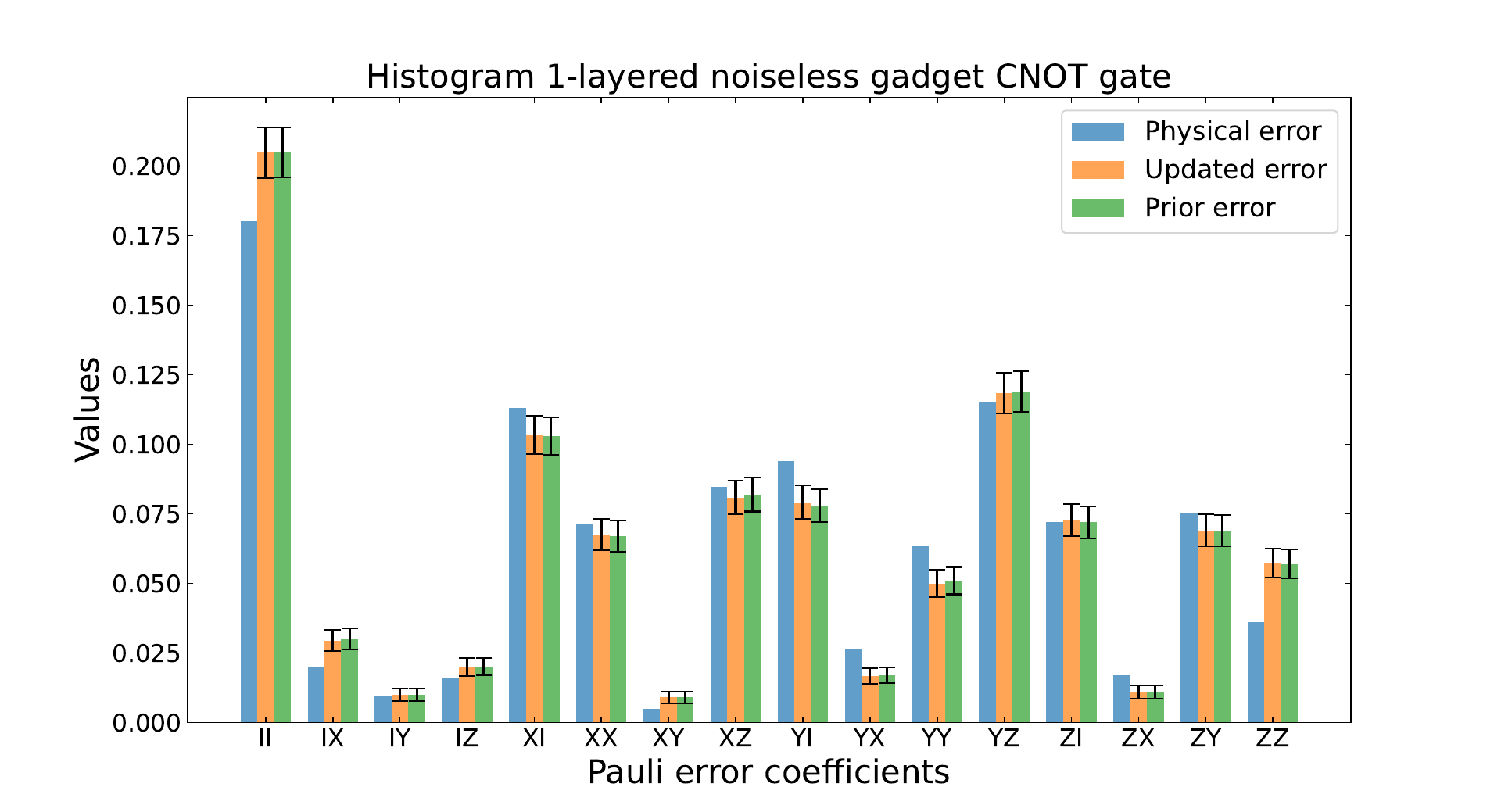}
    \caption{Histogram of the Pauli error estimates associated to a CNOT gate using noiseless $1$-layered gadgets. The `prior error' and the `updated error' columns display the expected values under the the prior distribution $\langle \lambda_i\rangle_{P^{(0)}}$ and the updated distribution $\langle \lambda_i\rangle_{P^{(n)}}$, respectively. The `physical error' column represents the error rates $\langle \lambda_i\rangle_{P_{\text{phy}}}$ of the real Pauli channel $\tilde{\Lambda}$ associated to the CNOT gate. To simulate device drift, the `physical error' values slightly deviates from the `prior error' estimates, hence by utilizing the flag gadgets we aim to learn their differences in real-time. The `updated error' columns where computed using the approximation (\ref{eq:approx_second_ord}), considering the set up of Fig.~\ref{fig:gadgets_repeated} starting from the $|00\rangle$ state, taking $n=1000$ repetitions and $\alpha_0=2000$. In addition, the numerical simulations were done using the stabilizer circuit simulator package stim \cite{gidney2021stim}. The $1$-layered flag gadgets were chosen uniformly at random from the $2$-qubit Pauli group excluding the identity $\mathcal{P}_2/\{I \otimes I\}$ (see Appendix~\ref{app:cnot_singe_lay} for further details).}
    \label{fig:approx_second_order}
\end{figure*}

After a series of $n$ consecutive flag gadget measurements, computing the expected values by numerically brute forcing the update rule (\ref{eq:update_general_expected}), using the closed form result (\ref{eq:beta_expectec}), is out of reach even for $n\gtrsim 10$. Thus, we opted to study an approximation to the exact update rule by considering an expansion in large $\alpha_0$, the only tunable parameter in our setup besides $n$ and the prior Pauli estimates. The rationale behind this choice is that the variance of a Dirichlet distribution is order $\text{Var}(\lambda_i)\lesssim 10^{-4}$ when $\alpha_0 \sim 2000$. Larger values of $\alpha_0$ assign more confidence to the prior distribution and require many shots $n$ to reliably update the expected values. On the other hand, smaller values of $\alpha_0$ assign less confidence to the prior distribution and the update rule is not robust enough -- for instance, the Bayesian update can be overly aggressive and modify error rates, even in the situation when the physical error model $\tilde{\Lambda}$ is similar to the prior estimate (small drift). Furthermore, we will show that the expansion in powers of $\alpha_0^{-1}$ showcases a peculiar structure that is hidden in the general formula (\ref{eq:update_general_expected}).

The expression for the expected values after $n$ shots up to linear order in large $\alpha_0$ is given by (see Appendix~\ref{app:alpha0} for a proof),
\begin{equation}
\begin{split}
    \langle \lambda_j \rangle_{P^{(n)}}&=\frac{\alpha_0}{\alpha_0+n}\langle \lambda_j \rangle_{P^{(0)}}\\
    +\frac{1}{\alpha_0+n}&\sum_{t=1}^n\frac{X_1(t)+\alpha_0^{-1}X_2(t)+\Omega(\alpha_0^{-2}n^3)}{1+\alpha_0^{-1}Y_2+\Omega(\alpha_0^{-2}n^3)}\:,
\label{eq:approx_second_ord}
\end{split}
\end{equation}
where the quantities $X_1(t)$, $X_2(t)$ and $Y_2$ are
\begin{equation}
    X_1(t)=\langle \lambda_j \rangle \frac{\delta(j,[g_t])}{(\sum_{i_t} \langle \lambda_{i_t}\rangle)},
\label{eq:x1}
\end{equation}
\begin{equation}
\begin{split}
    X_2(t)&=\langle \lambda_j \rangle \frac{\delta(j,[g_t])}{(\sum_{i_t} \langle \lambda_{i_t}\rangle)}\sum_{k=1,k\neq t}^{n}\Bigg(\frac{\delta(j,[g_k])}{(\sum_{i_k} \langle \lambda_{i_k}\rangle)}\\
    &\quad\quad\quad\quad+\sum_{q=1,q\neq t}^{k-1}\frac{(\sum_{i \in [g_k]\cap [g_q]}\langle \lambda_i \rangle)}{(\sum_{i_k} \langle \lambda_{i_k}\rangle)(\sum_{i_q} \langle \lambda_{i_q}\rangle)}\Bigg),
\label{eq:x2}
\end{split}
\end{equation}
\begin{equation}
    Y_2=\langle \lambda_j \rangle \sum_{k=1}^{n}\sum_{q=1}^{k-1}\frac{(\sum_{i \in [g_k]\cap [g_q]}\langle \lambda_i \rangle)}{(\sum_{i_k} \langle \lambda_{i_k}\rangle)(\sum_{i_q} \langle \lambda_{i_q}\rangle)}\:.
\label{eq:y2}
\end{equation}
The expected values in (\ref{eq:x1}), (\ref{eq:x2}) and (\ref{eq:y2}) should be understood as $\langle \lambda_j \rangle \equiv \langle \lambda_j \rangle_{P^{(0)}}$; also, a sum over $i_k$ denotes a sum over $i_k \in [g_k]$. Notice that further terms in (\ref{eq:approx_second_ord}) scale with higher powers of $\alpha_0n$, as emphasized by the Big-$\Omega$ notation. Thus, the expansion is only valid as long as $n < \alpha_0 $ because it becomes strongly coupled when $n\gtrsim \alpha_0$. The expected values (\ref{eq:approx_second_ord}) are guaranteed to be normalized since
\begin{equation}
    \sum_j \sum_{t=1}^n X_1(t)=n \quad , \quad \sum_j \sum_{t=1}^n X_2(t)=n Y_2\:.
\end{equation}
There are two important takeaways from (\ref{eq:approx_second_ord}). First, the expansion suggests that the exact update rule (\ref{eq:update_general_expected}) cannot be recursive. This is due to the existence of correlations between different iterations, manifested in the presence of terms that depend on $[g_k]\cap [g_q]$ with $k \neq q$ and $k,q=1,\cdots,n$. The root of these correlations can be traced-back to the derivation of the approximation, particularly, when expanding high-moments of the Dirichlet distribution (\ref{eq:beta_expectec}) in powers of $\alpha_0$ (see Eqn.~(\ref{eq:higher_mom_Dir_app}) in the Appendix). More generally, we expect higher-orders of $\alpha_0^{-p}$ to exhibit correlations between $p+1$ different iterations. 

Second, the expansion (\ref{eq:approx_second_ord}) truncated to the zero-th order in $\alpha_0^{-1}$ displays a tight connection with the maximal layered case (\ref{eq:expected_4gadgets}) and the $l$-layered single-shot $n=1$ case (\ref{eq:lmbda_1}). Concretely, those results emerge as particular cases of the zero-th order expansion, which has the following expression,
\begin{equation}
\begin{split}
    \langle \lambda_j \rangle_{P^{(n)}}&=\frac{\alpha_0 }{\alpha_0+n}\langle \lambda_j \rangle_{P^{(0)}}\\
    &\quad +\frac{\langle \lambda_j \rangle_{P^{(0)}}}{\alpha_0+n}\sum_{t=1}^{n}\frac{\delta(j,[g_t])}{(\sum_{i_t \in [g_t]}\langle \lambda_{i_t}\rangle_{P^{(0)}})}\:.
\label{eq:expected_order0}
\end{split}
\end{equation}
The result (\ref{eq:expected_order0}) is indeed recursive since
\begin{equation}
\begin{split}
    \langle \lambda_j\rangle_{P^{(n)}}&=\frac{\alpha_0+n-1}{\alpha_0+n} \langle \lambda_j\rangle_{P^{(n-1)}}\\
    &\quad+\frac{\langle \lambda_{j}\rangle_{P^{(0)}}}{\alpha_0+n}\frac{\delta(j,[g_n])}{\big(\sum_{i_n\in [g_n]} \langle \lambda_{i_n}\rangle_{P^{(0)}}\big)}\:.
\end{split}
\end{equation}

A complementary interpretation of (\ref{eq:expected_order0}) is that the updated probability distribution $P^{(n)}(\vec{\lambda})$ is also a Dirichlet distribution, whose coefficients $\vec{\alpha}$ can be computed from $\{\langle \lambda_i\rangle_{P^{(n)}}\}$ and the effective parameter $\alpha_0^{\text{eff}}=\alpha_0+n$. Using the same language as in (\ref{eq:bayes_single}), if at a given iteration the set $[g]$ is obtained from the measurement outcome $m_X$, the update rule that transforms the prior Dirichlet distribution  $D(\vec{\lambda},\vec{\alpha})$ into the updated Dirichlet distribution $D(\vec{\lambda},\vec{\alpha'})$, is a generalization of (\ref{eq:bayes_max_deltaf}) when $|[g]|>1$,
\begin{equation}
\alpha_i'=\alpha_i+\delta(i,[g])\frac{\langle \lambda_i \rangle_{D(\vec{\lambda},\vec{\alpha})}}{\big(\sum_{j \in [g]} \langle \lambda_j \rangle_{D(\vec{\lambda},\vec{\alpha})}\big)}\:.
\end{equation}

Finally, a numerical simulation of the first order expansion (\ref{eq:approx_second_ord}) for the most relevant practical case, with $l=1$ layers, can be found in Fig.~\ref{fig:approx_second_order}. Unfortunately, the performance of the noise channel adaptation is not good in this case. There are two issues at play. Firstly, due to the computational intractability of the exact update rule the approximate update rule must be utilized, and this leads to inaccuracies. Secondly, a single gadget does not provide enough information to isolate the error rates and thus the Bayesian updates are not effective. We conclude that the number of gadgets should be as close to the maximal $2n_q$ number as possible to enable effective noise adaptation.

\subsection{Impact of noisy gadgets}
\label{sec:noisyGadgets}
To model the impact of noise introduced by the flag gadgets, we assume that every part of them is faulty: including state preparation, gate implementation and measurements, see Fig.~\ref{fig:errors_general}. The precise definition of the error channels used in the numerical simulations is described with details in Appendix~\ref{app:num_methods}.

\begin{figure*}[ht!]
    \[ \quad \:\:\:\:
    \begin{array}{c}
        \Qcircuit @C=0.32em @R=.4em {
                     &\qw                     &\multigate{3}{L_3}&\multigate{6}{\Lambda_{L_3}}&\multigate{3}{L_2}&\multigate{5}{\Lambda_{L_2}}&\multigate{3}{L_1}&\multigate{4}{\Lambda_{L_1}}&\multigate{3}{U}&\multigate{3}{\Lambda}&\multigate{3}{R_1}&\multigate{4}{\Lambda_{R_1}}&\multigate{3}{R_2}&\multigate{5}{\Lambda_{R_2}}&\multigate{3}{R_3}&\multigate{6}{\Lambda_{R_3}}&\qw                     &\qw&\\
                     &\qw                     &\ghost{L_3}       &\ghost{\Lambda_{L_3}}       &\ghost{L_2}       &\ghost{\Lambda_{L_2}}       &\ghost{L_1}       &\ghost{\Lambda_{L_1}}       &\ghost{U}       &\ghost{\Lambda}       &\ghost{R_1}       &\ghost{\Lambda_{R_1}}       &\ghost{R_2}       &\ghost{\Lambda_{R_2}}       &\ghost{R_3}       &\ghost{\Lambda_{R_3}}       &\qw                     &\qw&\\
                     &                        &\nghost{L_3}      &\nghost{\Lambda_{L_3}}      &\nghost{L_2}      &\nghost{\Lambda_{L_2}}      &\nghost{L_1}      &\nghost{\Lambda_{L_1}}      &\nghost{U}      &\nghost{\Lambda}      &\nghost{R_1}      &\nghost{\Lambda_{R_1}}      &\nghost{R_2}      &\nghost{\Lambda_{R_2}}      &\nghost{R_3}      &\nghost{\Lambda_{R_3}}      &                        &\cdots &\\
                     &\qw                     &\ghost{L_3}       &\ghost{\Lambda_{L_3}}       &\ghost{L_2}       &\ghost{\Lambda_{L_2}}       &\ghost{L_1}       &\ghost{\Lambda_{L_1}}       &\ghost{U}       &\ghost{\Lambda}       &\ghost{R_1}       &\ghost{\Lambda_{R_1}}       &\ghost{R_2}       &\ghost{\Lambda_{R_2}}       &\ghost{R_2}       &\ghost{\Lambda_{R_3}}       &\qw                     &\qw&\\
  \lstick{|+\rangle} &\multigate{2}{\Lambda_I}&\qw               &\ghost{\Lambda_{L_3}}       &\qw               &\ghost{\Lambda_{L_2}}       &\ctrl{-1}         &\ghost{\Lambda_{L_1}}       &\qw             &\qw                   &\ctrl{-1}         &\ghost{\Lambda_{R_1}}       &\qw               &\ghost{\Lambda_{R_2}}       &\qw               &\ghost{\Lambda_{R_3}}       &\multigate{2}{\Lambda_M}&\measuretab{M_X}  \\
  \lstick{|+\rangle} &\ghost{\Lambda_I}       &\qw               &\ghost{\Lambda_{L_3}}       &\ctrl{-2}         &\ghost{\Lambda_{L_2}}       &\qw               &\qw                         &\qw             &\qw                   &\qw               &\qw                         &\ctrl{-2}         &\ghost{\Lambda_{R_2}}       &\qw               &\ghost{\Lambda_{R_3}}       &\ghost{\Lambda_M}       &\measuretab{M_X}  \\
  \lstick{|+\rangle} &\ghost{\Lambda_I}       &\ctrl{-3}         &\ghost{\Lambda_{L_3}}       &\qw               &\qw                         &\qw               &\qw                         &\qw             &\qw                   &\qw               &\qw                         &\qw               &\qw                         &\ctrl{-3}         &\ghost{\Lambda_{R_3}}       &\ghost{\Lambda_M}       &\measuretab{M_X} \\
        }
    \end{array}
    \]
    \caption{Example of a multi-layered flag gadget acting on a noisy unitary gate including quantum error channels for every noisy component. The quantum channels represent noise due to: the unitary gate $U$ ($\Lambda$), state initialization ($\Lambda_{I}$), measurement ($\Lambda_{M}$), and the controlled unitaries ($\Lambda_{L_i}$, $\Lambda_{R_i}$ for $i=1,\dots,l$).  In this example, there are $l=3$ layers.}   \label{fig:errors_general}
\end{figure*}

As a practical example, in Fig.~\ref{fig:4layers_CNOT_noisy} we show a numerical simulation of $4$-layered noisy gadgets acting on a CNOT gate. The result of Fig.~\ref{fig:4layers_CNOT_noisy} has to be contrasted with Fig.~\ref{fig:4layers_CNOT}, where the flag gadgets are assumed to be noiseless. We can observe that both histograms give similar updated error channels for a low physical error rate.

\begin{figure*}[ht]
    \centering
    \includegraphics[width=0.8\textwidth]{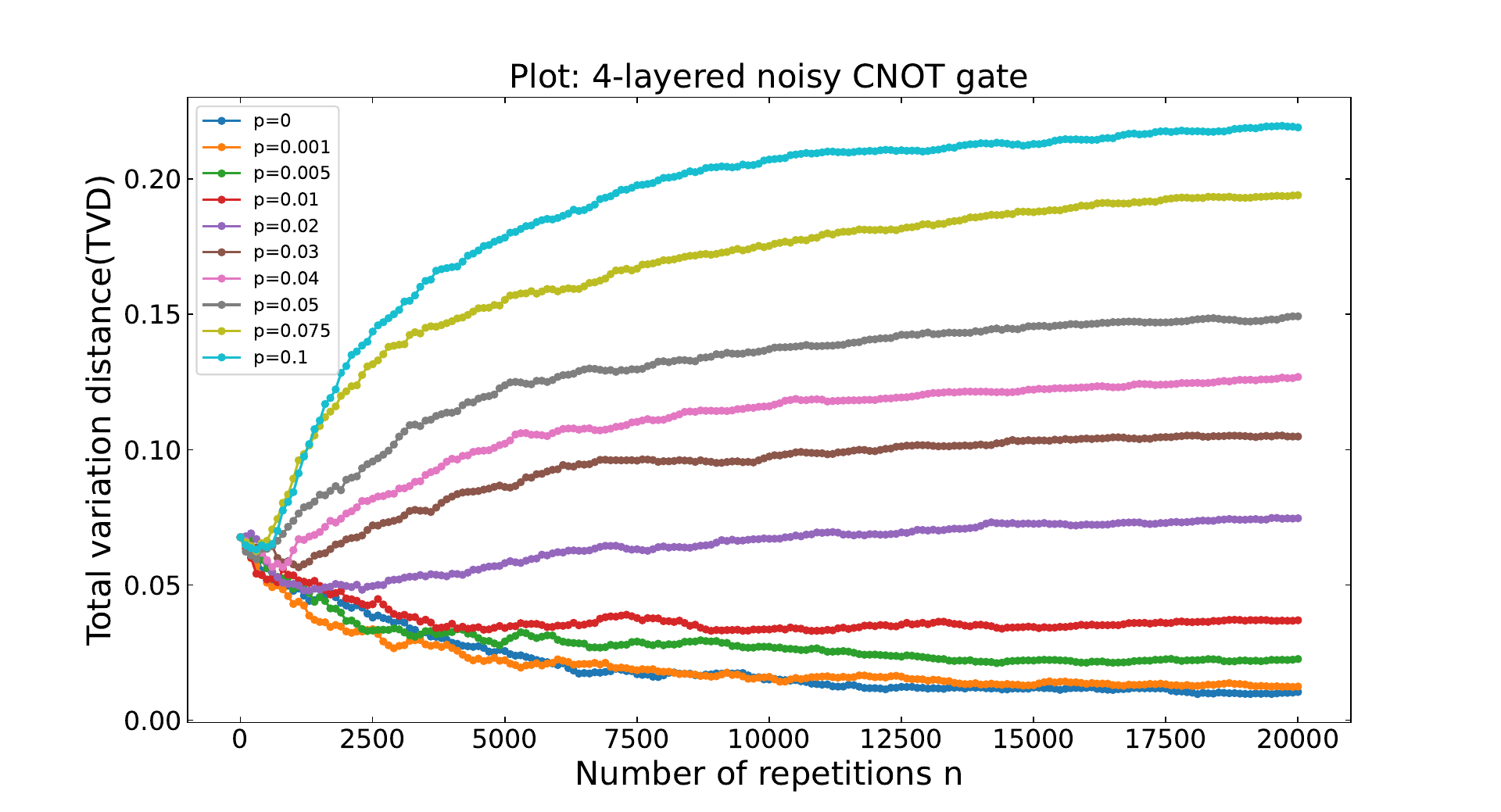}
    \caption{Total variation distance (TVD) (\ref{eq:distance}) between Pauli error rates updates $\langle \lambda_j\rangle_{P^{(n)}}$ (\ref{eq:expected_4gadgets}) and physical error rates $\langle \lambda_j\rangle_{P_{\text{phy}}}$ (see Fig.~\ref{fig:4layers_CNOT}) as a function of the number of repetitions $n$ for $4$-layered flag gadgets acting on a CNOT gate. The numerical simulations were done considering the set up of Fig.~\ref{fig:gadgets_repeated} starting from the $|00\rangle$ state, taking $n\in [1,20000]$ repetitions with $\alpha_0=2000$, and using the stabilizer circuit simulator package stim \cite{gidney2021stim}. The noisy flag gadgets were modeled using a uniform depolarizing error model of strength $p$ (see Section~\ref{sec:noisyGadgets}).}
    \label{fig:noisy_4layers_vsp}
\end{figure*}

To quantify the effect of the noise introduced by the flag gadgets, we utilize the total variation distance (TVD) $d_{\text{TVD}}$ as a figure of merit:
\begin{equation}
    d_{\text{TVD}}=\frac{1}{2}\sum_{i=1}^{4^{n_q}} \Big|\langle \lambda_i\rangle_{P^{(n)}}-\langle \lambda_i\rangle_{P_{\text{phy}}}\Big|\:,
\label{eq:distance}
\end{equation}
which compares the updated Pauli error rates after $n$ steps $\{\langle \lambda_j\rangle_{P^{(n)}}\}$ against the actual physical error rates $\{\langle \lambda_j\rangle_{P_{\text{phy}}}\}$. A numerical simulation of the TVD as a function of $n$, for several noise strength values $p$, can be found in Fig.~\ref{fig:noisy_4layers_vsp}. Given the set of physical error rates employed in the numerical simulations, we observe that when $p \gtrsim 0.02$ the TVD does not decay for large $n$. Moreover, after sufficient repetitions the TVD becomes bigger than the distance with respect to the prior Pauli estimates. This observation suggests that the Bayesian inference is not robust to noise in the flag gadgets above a certain threshold.

It is possible that the Bayesian update rule can be modified to incorporate the noise due to flag gadgets if the latter is well characterized. Here we show how this can be done for \textit{measurement noise}. For example, consider a probability of $X$-type measurement error of $p$.
In such a case, the likelihood probability in the Bayesian analysis (Eqn. \eqref{eq:likelihood_orig})
should be amended as
\begin{equation}
\begin{split}    P(m_X|\vec{\lambda})&=P(m_X|\vec{\lambda};\text{0-errors})\tilde{P}(m_X;\text{0-errors})\\
&\:  + \sum_{k=1}^l P(m_X|\vec{\lambda};\text{$k$-errors})\tilde{P}(m_X;\text{$k$-errors}),
\end{split}
\end{equation}
where $\tilde{P}(m_X;\text{$k$-errors})$ indicates the probability of having $k$ measurement errors and $P(m_X|\vec{\lambda};\text{$k$-errors})$ indicates the likelihood probability with $k$ measurement errors.

In case that every measurement error occurs with probability $p$, then the likelihood probability for the \textit{maximal number of layers} case ($l=2n_q$) is
\begin{equation}
\begin{split}
    P(m_X|\vec{\lambda})&=(1-p)^{2n_q}\lambda_i\\
    &\quad + \sum_{k=1}^{2n_q} p^k (1-p)^{2n_q-k}\left(\sum_{j/\text{wt}(P_i,P_j)=k}\lambda_j \right)\:.
\label{eq:likelihood
_maximal}
\end{split}
\end{equation}
Here $\binom{2n_q}{k}$ is the binomial coefficient, and $\text{wt}(P_i,P_j)$ is the Hamming distance between two Pauli matrices $P_i$ and $P_j$. The later one is defined as the number of different entries in the binary representation of $P_i$\footnote{The Pauli matrix $P_i$ stands for the matrix singled out by the measurement outcome $m_X$ of the maximal layered flag gadget in a noiseless scenario, i.e., $i \in [g]$ (see Section~\ref{sec:maxiam_case}).} and $P_j$. Recall that the binary representation of a Pauli matrix $Q$ acting on $n_q$ qubits is given by the $2n_q$ dimensional string $(c_1,\cdots,c_{n_q},d_1,\cdots,d_{n_q})$ such that
\begin{equation}
    Q \sim \bigotimes_{s=1}^{n_q} X_s^{c_s}Z_s^{d_s}\quad , \quad c_s,d_s \in \{0,1\}\:,
\label{eq:pauliQ_bin}
\end{equation}
where we have ignored the global phase. Note that given $P_i$, there are exactly $\binom{2n_q}{k}$ matrices $P_j$ such that $\text{wt}(P_i,P_j)=k$. The proof of formula (\ref{eq:likelihood
_maximal}) has two parts: i) since single qubit measurement errors occur with probability $p$, the probability of not measuring any error is just $(1-p)^{2n_q}$. In addition, the factor $\lambda_i$ is associated with the Pauli matrix $P_i$ singled out by the noiseless measurement outcome $m_X$, i.e., with $i \in [g]$; ii) if $k$ error occurs, they will appear with probability $p^k (1-p)^{2n_q-k}$. The coefficients $\lambda_j$ that are singled out have Hamming distance $\text{wt}(P_i,P_j)=k$ because of the structure of the flag gadgets for the maximal layered case. This is because the right unitaries $R_1,\cdots ,R_{2n_q}$ are single qubit $X$ and $Z$ gates acting on every data qubit. The commutation/anti-commutation relations of a given Pauli $Q$ (\ref{eq:pauliQ_bin}) with every single qubit gate $X_s(Z_s)$ acting on the $s$-th qubit with $s=1,\dots,n_q$ are 
\begin{equation}
    [Q,X_s]_{b_q}=0 \: \Longleftrightarrow \:d_s =\sin\left(\frac{b_q+3}{4}\pi\right)\:,
\label{eq:com_X}
\end{equation}
\begin{equation}
    [Q,Z_s]_{b_q}=0 \: \Longleftrightarrow \:c_s = \sin\left(\frac{b_q+3}{4}\pi\right) \:.
\label{eq:com_Z}
\end{equation}
Recall that $b_q=\pm 1$ (see (\ref{eq:com_ant})) is given by the measurement outcome of each layer $b_q=m_X^{(q)}$, with $q=1,\dots,2n_q$ (\ref{eq:set_g_l}).
Hence, given an initial matrix $P_i$, all the commutators/anti-commutators with single qubit gates $X_s(Z_s)$ that  are equal to zero are known. In case that exactly $k$ errors occur, only $k$ of them will be different with respect to their original value (in other words, only $k$ generalized commutators will flip the value $b_q$ of $[P_i,X_s]_{b_q}$) and $2n_q-k$ will be unaltered. Since Pauli matrices either commute or anti-commute, equations (\ref{eq:com_X}) and (\ref{eq:com_Z}) then suggest that this new condition could be achieved by only changing $k$ different elements of the binary representation of the Pauli $P_i$. This indeed corresponds to the aforementioned Hamming distance condition for $P_j$.

In contrast, the modified likelihood probability for the \textit{single-layered} case ($l=1$) has a simpler form
\begin{equation}
\begin{split}    P(m_X|\vec{\lambda})=(1-p)\Big(\sum_{i \in [g]}\lambda_{i}\Big)+p \Big(1-\sum_{i \in [g]}\lambda_{i}\Big).
\end{split}
\label{eq:likelihood_single}
\end{equation}
The proof of (\ref{eq:likelihood_single}) follows a similar logic as the proof of (\ref{eq:likelihood
_maximal}). In either case, both expressions (\ref{eq:likelihood
_maximal}) and (\ref{eq:likelihood_single}) contain more than a single $\lambda_i$ in their respective sums. Hence, as argued in Section~\ref{sec:genraliz_update}, after several iterations of the flag gadgets, the probability distribution will be a convex combination of Dirichlet distributions that grows exponentially, so in practice some sort of approximation has to be implemented. We postpone this kind of analysis for future work.

To conclude this section, we would like to show explicit formulas for the update rules in the specific case of noisy single-layered flag gadgets. After a single iteration, Bayes' equation becomes
\begin{equation}
    P^{(1)}_{\text{noisy}}=\frac{p+(1-2p)\left(\sum_{i_1 \in [g_1]}\lambda_{i_1}\right)}{p+(1-2p)\left(\sum_{i_1 \in [g_1]}\langle\lambda_{i_1}\rangle_{P^{(0)}}\right)}P^{(0)}\:.
\end{equation}
Thus, the update rule for the expected values are
\begin{equation}
\begin{split}
    \langle \lambda_j\rangle_{P^{(1)}}^{\text{noisy}}&=\frac{p}{p+(1-2p)\left(\sum_{i_1 \in [g_1]}\langle\lambda_{i_1}\rangle_{P^{(0)}}\right)}\langle \lambda_j\rangle_{P^{(0)}}\\&+\frac{(1-2p)\left(\sum_{i_1 \in [g_1]}\langle\lambda_{i_1}\rangle_{P^{(0)}}\right)}{p+(1-2p)\left(\sum_{i_1 \in [g_1]}\langle\lambda_{i_1}\rangle_{P^{(0)}}\right)}\langle \lambda_j \rangle_{P^{(1)}}\:,
\end{split}
\end{equation}
where $\langle \lambda_j \rangle_{P^{(1)}}$ is defined as in (\ref{eq:lmbda_1}). Intuitively, for $p>0$, the first terms slows down the updating of the error rates, since the measurement results cannot be fully trusted due to measurement errors.

\section{Conclusions}
\label{sec:conclusions}
We have developed and analyzed a technique for adapting the estimates of noise channels associated to gates on a quantum computer. This is a task that is critical for the success of modern error mitigation techniques because time-dependent fluctuations of device parameters (drift) makes estimates of noise channels stale within the execution time of quantum algorithms. The technique uses \emph{extended flag gadgets} to gather information about gate noise without interrupting the execution of the underlying algorithmic circuit. We show how the information gathered from the flag gadget measurements can be used within a Bayesian inference framework to update parameters associated to a noise channel. 

Through numerical simulations we showed that the technique is very effective in the regime where a the maximum number of extended flag gadgets ($2n_q$ for a gate operating on $n_q$ qubits) is used. This adds at least $4n_q$ additional controlled gates to ancilla qubits, and thus it is desirable to also study the performance when fewer extended flag gadgets are used. However, our numerical studies show that the technique rapidly becomes less effective as the number of flag gadgets is decreased, essentially due to a mismatch between the small amount of information available from measurements (1 bit per flag gadget) and the large number of parameters that must be updated.  In addition, our numerical studies show that noise on the gates implementing the extended flag gadgets corrupts the information gain and slows down the convergence of the noise channel adaptation process. As a result, the technique we develop here is most useful in the low noise regime where one can afford to implement the maximum number of flag gadgets and the impact of noise on their implementation is minimal. 

\par\medskip\noindent 
\textbf{Acknowledgments.} We thank Isaac H. Kim and Stefan Seritan for useful discussions. 
This work was supported by the U.S. Department of Energy, Office of Science, Office of Advanced Scientific Computing Research through the Accelerated Research in Quantum Computing Program and the MACH-Q project. 

\bibliography{bib_flag_gadgets_paper}

\appendix
\onecolumngrid 
\newpage

\section{Flag gadgets for the CNOT gate}
\label{app:app_CNOT_all}
\subsection{$4$-layered flag gadgets}
\label{app:4layeredCNOT}
\begin{figure}[ht!]
\[
\begin{array}{c}
\Qcircuit @C=0.32em @R=.4em  {    
                   &\multigate{1}{ZZ}&\gate{Z} &\qw      &\multigate{1}{XX}&\ctrl{+1}&\multigate{1}{\tilde{\Lambda}}&\gate{X} &\qw      &\gate{Z} &\qw      &\qw      &      \\
                   &\ghost{ZZ}       &\qw      &\gate{X} &\ghost{XX}       &\targ    &\ghost{\tilde{\Lambda}}       &\qw      &\gate{X} &\qw      &\gate{Z} &\qw      &      \\
\lstick{|+\rangle} &\qw              &\qw      &\qw      &\ctrl{-1}        &\qw      &\qw                           &\ctrl{-2}&\qw      &\qw      &\qw      & \measuretab{M_X} &\:\:\:\:\:\:\:\:\:\: m_X^{(1)} \\
\lstick{|+\rangle} &\qw              &\qw      &\ctrl{-2}&\qw              &\qw      &\qw                           &\qw      &\ctrl{-2}&\qw      &\qw      & \measuretab{M_X} &\:\:\:\:\:\:\:\:\:\: m_X^{(2)}\\
\lstick{|+\rangle} &\qw              &\ctrl{-4}&\qw      &\qw              &\qw      &\qw                           &\qw      &\qw      &\ctrl{-4}&\qw      & \measuretab{M_X} &\:\:\:\:\:\:\:\:\:\: m_X^{(3)}\\
\lstick{|+\rangle} &\ctrl{-4}        &\qw      &\qw      &\qw              &\qw      &\qw                           &\qw      &\qw      &\qw      &\ctrl{-4}& \measuretab{M_X} &\:\:\:\:\:\:\:\:\:\: m_X^{(4)} \gategroup{1}{6}{2}{7}{.3em}{--}\\
}  
\end{array}
\]
\caption{$4$-layered gadget for a noisy CNOT gate. The dashed region represents a noisy CNOT gate, modeles as noiseless CNOT gate followed by a Pauli channel $\tilde{\Lambda}$.} \label{fig:2qubitGadget} \end{figure}
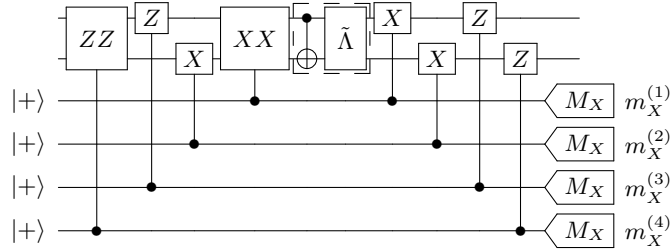

In this Appendix we provide details regarding the $4$-layered flag gadgets utilized for updating the noise channel estimates associated to a CNOT gate. In Fig.~\ref{fig:2qubitGadget}, we show the explicit construction of the flag gadgets. Note that single-controlled multi-qubit gates can be decomposed as a sequence of single-controlled single-qubit gates. For the case in question, $C_3X_1X_2=C_3X_1 \circ C_3X_2$ and $C_3Z_1Z_2=C_3Z_1 \circ C_3Z_2$.

In Table~\ref{tab:likelihood_cnot}, we show the likelihood probability (\ref{eq:likelihood_orig}) which depends on the measurement outcome $m_X$ of Fig.~\ref{fig:2qubitGadget}. 

\begin{table}[!ht]
\centering
\begin{tabular}{|c|c|c|c|c|}
\hline
$m_X^{(1)}$ & $m_X^{(2)}$ & $m_X^{(3)}$ & $m_X^{(4)}$ & $P(m_X|\vec{\lambda})$ \\ \hline
$+1$        & $+1$        & $+1$        & $+1$        & $\lambda_{II}$    \\ \hline
$+1$        & $+1$        & $+1$        & $-1$        & $\lambda_{IX}$    \\ \hline
$+1$        & $+1$        & $-1$        & $+1$        & $\lambda_{XI}$    \\ \hline
$+1$        & $-1$        & $+1$        & $+1$        & $\lambda_{IZ}$    \\ \hline
$-1$        & $+1$        & $+1$        & $+1$        & $\lambda_{ZI}$    \\ \hline
$+1$        & $+1$        & $-1$        & $-1$        & $\lambda_{XX}$    \\ \hline
$+1$        & $-1$        & $+1$        & $-1$        & $\lambda_{IY}$    \\ \hline
$-1$        & $+1$        & $+1$        & $-1$        & $\lambda_{ZX}$    \\ \hline
$+1$        & $-1$        & $-1$        & $+1$        & $\lambda_{XZ}$    \\ \hline
$-1$        & $+1$        & $-1$        & $+1$        & $\lambda_{YI}$    \\ \hline
$-1$        & $-1$        & $+1$        & $+1$        & $\lambda_{ZZ}$    \\ \hline
$+1$        & $-1$        & $-1$        & $-1$        & $\lambda_{XY}$    \\ \hline
$-1$        & $+1$        & $-1$        & $-1$        & $\lambda_{YX}$    \\ \hline
$-1$        & $-1$        & $+1$        & $-1$        & $\lambda_{ZY}$    \\ \hline
$-1$        & $-1$        & $-1$        & $+1$        & $\lambda_{YZ}$    \\ \hline
$-1$        & $-1$        & $-1$        & $-1$        & $\lambda_{YY}$    \\ \hline
\end{tabular}
\caption{Likelihood probability of measuring $m_X$ given a prior probability distribution $P(\vec{\lambda})$ for a $4$-layered flag gadget acting on a $2$-qubit unitary. The right column was constructed using (\ref{eq:set_g_l}), and indices $i\in [1,16]$ where replaced by $i \in \{I,X,Y,Z\}^{\otimes 2}$, respectively. This table represents the $2$-qubit generalization of Table~\ref{tab:likelihood_singleq}.}
\label{tab:likelihood_cnot}
\end{table}

\subsection{$1$-layered flag gadgets}
\label{app:cnot_singe_lay}
In this Appendix we provide details regarding the $1$-layered flag gadgets utilized for updating the noise channel estimates associated to a CNOT gate. As it was mentioned in Fig.~\ref{fig:approx_second_order}, the $1$-layered flag gadgets were chosen uniformly at random from the $2$-qubit Pauli group excluding the identity $\mathcal{P}_2/\{I \otimes I\}$. The set $[g]$ (\ref{eq:set_g_l}) associated to with every gadget $R \in \mathcal{P}_2/\{I \otimes I\}$ is described in Table~\ref{tab:single_lay_gadgets}. We opted to use the convention of Table~\ref{tab:2qubit_label} as a label for the indices of $\lambda_i$.

\begin{table}[!ht]
\centering
\begin{tabular}{|c|c|c|c|}
\hline
$R$ & $L$ & $[g]:\: m_X=+1$ & $[g]:\: m_X=-1$  \\ \hline
$X_2$ & $X_2$ & $[1, 2, 5, 6, 9, 10, 13, 14]$ & $[3, 4, 7, 8, 11, 12, 15, 16]$\\
$Y_2$ & $Z_1Y_2$ & $[1, 3, 5, 7, 9, 11, 13, 15]$ & $[2, 4, 6, 8, 10, 12, 14, 16]$\\
$Z_2$ & $Z_1Z_2$ & $[1, 4, 5, 8, 9, 12, 13, 16]$ & $[2, 3, 6, 7, 10, 11, 14, 15]$\\
$X_1$ & $X_1X_2$ & $[1, 2, 3, 4, 5, 6, 7, 8]$ & $[9, 10, 11, 12, 13, 14, 15, 16]$\\
$Y_1$ & $Y_1X_2$ & $[1, 2, 3, 4, 9, 10, 11, 12]$ & $[5, 6, 7, 8, 13, 14, 15, 16]$\\
$Z_1$ & $Z_1$ & $[1, 2, 3, 4, 13, 14, 15, 16]$ & $[5, 6, 7, 8, 9, 10, 11, 12]$\\
$X_1X_2$ & $X_1$ & $[1, 2, 5, 6, 11, 12, 15, 16]$ & $[3, 4, 7, 8, 9, 10, 13, 14]$\\
$X_1Y_2$ & $Y_1Z_2$ & $[1, 3, 5, 7, 10, 12, 14, 16]$ & $[2, 4, 6, 8, 9, 11, 13, 15]$\\
$X_1Z_2$ & $-Y_1Y_2$ & $[1, 4, 5, 8, 10, 11, 14, 15]$ & $[2, 3, 6, 7, 9, 12, 13, 16]$\\
$Y_1X_2$ & $Y_1$ & $[1, 2, 7, 8, 9, 10, 15, 16]$ & $[3, 4, 5, 6, 11, 12, 13, 14]$\\
$Y_1Y_2$ & $-X_1Z_2$ & $[1, 3, 6, 8, 9, 11, 14, 16]$ & $[2, 4, 5, 7, 10, 12, 13, 15]$\\
$Y_1Z_2$ & $X_1Y_2$ & $[1, 4, 6, 7, 9, 12, 14, 15]$ & $[2, 3, 5, 8, 10, 11, 13, 16]$\\
$Z_1X_2$ & $Z_1X_2$ & $[1, 2, 7, 8, 11, 12, 13, 14]$ & $[3, 4, 5, 6, 9, 10, 15, 16]$\\
$Z_1Y_2$ & $Y_2$ & $[1, 3, 6, 8, 10, 12, 13, 15]$ & $[2, 4, 5, 7, 9, 11, 14, 16]$\\
$Z_1Z_2$ & $Z_2$ & $[1, 4, 6, 7, 10, 11, 13, 16]$ & $[2, 3, 5, 8, 9, 12, 14, 15]$\\
\hline
\end{tabular}
\caption{Set of $1$-layered flag gadgets for a unitary $U=\text{CNOT}_{1 \to 2}$ including: right unitaries $R \in \mathcal{P}_2/\{I \otimes I\}$, left unitaries satisfying $RUL=U$ (\ref{eq:equivalence}), and set of indices $[g]$ dependent on the measurement outcome $m_X$ (\ref{eq:set_g}). The indices conventions for the Pauli estimates $\lambda_i$ can be found in Fig.~\ref{tab:2qubit_label}.}
\label{tab:single_lay_gadgets}
\end{table}

\begin{figure}[!ht]
\[
\begin{array}{c}
\Qcircuit @C=.7em @R=.4em @! {
&\ctrl{1} &\qw &\ctrl{1} & \qw\\
&\targ  &\gate{Z} &\targ&\qw}
\end{array}=
\begin{array}{c}
\Qcircuit @C=.7em @R=.4em @! {
&\gate{Z}& \qw\\
&\gate{Z} &\qw}
\end{array}
\:\:\:\:\:,\:\:\:\:
\begin{array}{c}
\Qcircuit @C=.7em @R=.4em @! {
&\ctrl{1} &\gate{X} &\ctrl{1} & \qw\\
&\targ  &\qw &\targ&\qw}
\end{array}=
\begin{array}{c}
\Qcircuit @C=.7em @R=.4em @! {
&\gate{X}& \qw\\
&\gate{X} &\qw}
\end{array}\quad\:.
\]
\caption{Non-trivial gate identities for the conjugation of single qubit $X$($Z$) gates under CNOT gates.} \label{fig:prop_errors_Conj} 
\end{figure}
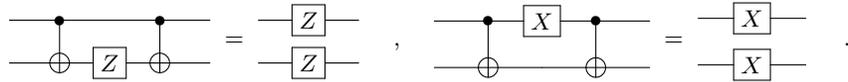

\begin{table}[!ht]
\centering
\begin{tabular}{|c|c|c|c|c|c|c|c|c|c|c|c|c|c|c|c|c|}
\hline
$\lambda_j$&$II$&$IX$&$IY$&$IZ$&$XI$&$XX$&$XY$&$XZ$&$YI$&$YX$&$YY$&$YZ$&$ZI$&$ZX$&$ZY$&$ZZ$\\ \hline
$j$    &$1$&$2$&$3$&$4$&$5$&$6$&$7$&$8$&$9$&$10$&$11$&$12$&$13$&$14$&$15$&$16$ \\
\hline
\end{tabular}
\caption{Labeling for Pauli error estimates $\lambda_i$ associated to $2$-qubit gates.}
\label{tab:2qubit_label}
\end{table}
To construct Table~\ref{tab:single_lay_gadgets} it was necessary to use the fundamental properties of Fig.~\ref{fig:prop_errors_Conj}, 


\section{Proof of the approximate update rule}
\label{app:alpha0}
In this Appendix we derive the approximate update rule (\ref{eq:approx_second_ord}) from the exact expression (\ref{eq:update_general_expected}), which we reproduce here again
\begin{equation}
    \langle \lambda_j \rangle_{P^{(n)}}=\frac{\sum_{i_n\in [g_n]} \ldots \sum_{i_1\in [g_1]}  \langle \lambda_{j}\lambda_{i_n}\ldots \lambda_{i_1}\rangle_{P^{(0)}}}{\sum_{i_n\in [g_n]} \ldots \sum_{i_1\in [g_1]}  \langle \lambda_{i_n}\ldots \lambda_{i_1}\rangle_{P^{(0)}}}\:.
\label{eq:app_gen_Expected}
\end{equation}
Given that product of high moments of a Dirichlet distribution have a close form expression (\ref{eq:beta_expectec}) in terms of the generalized beta function (\ref{eq:gen_beta_function}), we can employ the recursive property of the Gamma function $\Gamma(x+1)=x\Gamma(x)$ to express
\begin{equation}
     \langle \lambda_{j}\lambda_{i_n}\ldots \lambda_{i_1}\rangle_{P^{(0)}}= \langle \lambda_{i_n}\ldots \lambda_{i_1}\rangle_{P^{(0)}}\left(\frac{\alpha_0}{\alpha_0+n}\langle \lambda_j\rangle_{P^{(0)}}+\frac{1}{\alpha_0+n}(\delta_{j,i_1}+\ldots+\delta_{j,i_n})\right) \:.
\label{eq:recursive_expec_j}
\end{equation}
Replacing (\ref{eq:recursive_expec_j}) into (\ref{eq:app_gen_Expected}) gives
\begin{equation}
     \langle \lambda_j \rangle_{P^{(n)}}=\frac{\alpha_0}{\alpha_0+n}\langle \lambda_j \rangle_{P^{(0)}}+\frac{1}{\alpha_0+n}\sum_{t=1}^n \frac{\sum_{i_n}\dots\sum_{i_1}\langle \lambda_{i_n}\ldots \lambda_{i_1}\rangle_{P^{(0)}}\delta_{j,i_t}}{\sum_{i_n}\dots\sum_{i_1} \langle \lambda_{i_n}\ldots \lambda_{i_1}\rangle_{P^{(0)}}}\:.
\label{eq:expected_j_as_delta}
\end{equation}
Note that we introduced the notation that a sum over $i_k$ denotes a sum over $i_k \in [g_k]$. To proceed, we can expand out the correlators even further using $n-1$ times the recursive relation
\begin{equation}
\begin{split}
     \langle \lambda_{i_n}\ldots \lambda_{i_1}\rangle_{P^{(0)}}&=\prod_{k=1}^{n}\frac{1}{\alpha_0+k-1}\left(\alpha_0 \langle \lambda_{i_k}\rangle_{P^{(0)}}+\sum_{p=1}^{k-1}\delta_{i_k,i_p}\right)\\
     &=\frac{\Gamma(\alpha_0+n-1)}{\Gamma(\alpha_0+1)}\alpha_0^n \left(\prod_{p=1}^n \langle \lambda_{i_p}\rangle \right)\left(1+\alpha_0^{-1}\sum_{k=1}^n \sum_{q=1}^{k-1}\frac{\delta_{i_k,i_q}}{\langle \lambda_{i_k}\rangle}+\Omega(\alpha_0^{-2}n^3)\right)\:.
\label{eq:higher_mom_Dir_app}
\end{split}
\end{equation}
In the last step, we expanded out the numerator in powers of $\alpha_0$ and we also removed the $P^{(0)}$ label from the expected values. At order $\alpha_0^{-p}$ in (\ref{eq:higher_mom_Dir_app}) the individual terms have the product of exactly $p$ different Kronecker deltas. This expansion property explains the correlation between different iterations discussed in Section~\ref{sec:approximae_rule}. For instance, at order $\alpha_0^{-1}$, upon taking the respective sums there will be indices belonging to the intersection of sets from two different iterations, i.e., $[g_k]\cap [g_q]$ with $k\neq q$ and $k,q=1,\dots,n$.

We are now ready to replace (\ref{eq:higher_mom_Dir_app}) into (\ref{eq:expected_j_as_delta}). The non-trivial denominator of (\ref{eq:expected_j_as_delta}) can be then written as
\begin{equation}
\begin{split}
\sum_{i_n}\dots\sum_{i_1}\langle \lambda_{i_n}\ldots \lambda_{i_1}\rangle_{P^{(0)}}&=\frac{\Gamma(\alpha_0+n-1)}{\Gamma(\alpha_0+1)}\alpha_0^n \left[\prod_{p=1}^n \left( \sum_{i_p}\langle\lambda_{i_p}\rangle \right) \right]\Bigg(1+\\
&+ \alpha_0^{-1}\sum_{k=1}^{n}\sum_{q=1}^{k-1}\frac{(\sum_{i \in [g_k]\cap [g_q]}\langle \lambda_i \rangle)}{(\sum_{i_k} \langle \lambda_{i_k}\rangle)(\sum_{i_q} \langle \lambda_{i_q}\rangle)}+\Omega(\alpha_0^{-2}n^3)\Bigg)\:.
\label{eq:app_denom}
\end{split}
\end{equation}
The analysis of the non-trivial numerator of (\ref{eq:expected_j_as_delta}) is a bit more subtle, because of the additional delta function 
\begin{equation}
\begin{split}
\sum_{i_n}\dots\sum_{i_1}\langle \lambda_{i_n}\ldots \lambda_{i_1}\rangle_{P^{(0)}}\delta_{j,i_t}&=\frac{\Gamma(\alpha_0+n-1)}{\Gamma(\alpha_0+1)}\alpha_0^n \left[\prod_{p=1}^n \left( \sum_{i_p}\langle\lambda_{i_p}\rangle \right) \right]\langle \lambda_j\rangle\Bigg(\frac{\delta(j,[g_t])}{(\sum_{i_t} \langle \lambda_{i_t}\rangle)}+\\
&+ \alpha_0^{-1}\frac{\delta(j,[g_t])}{(\sum_{i_t} \langle \lambda_{i_t}\rangle)}\sum_{k=1,k\neq t}^{n}\sum_{q=1,q\neq t}^{k-1}\frac{(\sum_{i \in [g_k]\cap [g_q]}\langle \lambda_i \rangle)}{(\sum_{i_k} \langle \lambda_{i_k}\rangle)(\sum_{i_q} \langle \lambda_{i_q}\rangle)}\\
&+\alpha_0^{-1}\sum_{k=1,k\neq t}^n\frac{\delta(j,[g_k]\cap [g_t])}{(\sum_{i_k} \langle \lambda_{i_k}\rangle)(\sum_{i_t} \langle \lambda_{i_t}\rangle)}+\Omega(\alpha_0^{-2}n^3)\Bigg)\:.
\end{split}
\label{eq:app_num}
\end{equation}
Putting (\ref{eq:app_num}) and (\ref{eq:app_denom}) together into (\ref{eq:expected_j_as_delta}) provides a proof for the approximate update rule (\ref{eq:approx_second_ord})
\begin{equation}
\begin{split}
    \langle \lambda_j \rangle_{P^{(n)}}=\frac{\alpha_0}{\alpha_0+n}\langle \lambda_j \rangle_{P^{(0)}}+\frac{1}{\alpha_0+n}\sum_{t=1}^n\frac{X_1(t)+\alpha_0^{-1}X_2(t)+\Omega(\alpha_0^{-2}n^3)}{1+\alpha_0^{-1}Y_2+\Omega(\alpha_0^{-2}n^3)}\:.
\end{split}
\end{equation}
Where the quantities $X_1(t)$, $X_2(t)$ and $Y_2$ are
\begin{equation}
    X_1(t)=\langle \lambda_j \rangle \frac{\delta(j,[g_t])}{(\sum_{i_t} \langle \lambda_{i_t}\rangle)},
\end{equation}
\begin{equation}
    X_2(t)=\langle \lambda_j \rangle \frac{\delta(j,[g_t])}{(\sum_{i_t} \langle \lambda_{i_t}\rangle)}\sum_{k=1,k\neq l}^{n}\left(\frac{\delta(j,[g_k])}{(\sum_{i_k} \langle \lambda_{i_k}\rangle)}+\sum_{q=1,q\neq t}^{k-1}\frac{(\sum_{i \in [g_k]\cap [g_q]}\langle \lambda_i \rangle)}{(\sum_{i_k} \langle \lambda_{i_k}\rangle)(\sum_{i_q} \langle \lambda_{i_q}\rangle)}\right),
\end{equation}
\begin{equation}
    Y_2=\langle \lambda_j \rangle \sum_{k=1}^{n}\sum_{q=1}^{k-1}\frac{(\sum_{i \in [g_k]\cap [g_q]}\langle \lambda_i \rangle)}{(\sum_{i_k} \langle \lambda_{i_k}\rangle)(\sum_{i_q} \langle \lambda_{i_q}\rangle)}.
\end{equation}

\section{Numerical Methods}
\label{app:num_methods}

The numerical simulations are done using the stabilizer circuit simulator Stim~\cite{gidney2021stim} in Python. In Stim qubits are by default initialized in the computational basis $|0\rangle$ states. Clifford gates and single qubit destructive measurements are supported. Error models can be incorporated before or after the desired operations, and they are parametrized by some noise strength parameters, respectively. 
\newline 
\indent Regarding error models, for Figs. \ref{fig:4layers_CNOT},~\ref{fig:4layers_CNOT_noisy},~\ref{fig:approx_second_order} and~\ref{fig:noisy_4layers_vsp} the prior expected values $\langle \lambda_i\rangle_{P^{(0)}}$ are chosen at random from a $K=16$ variable Dirichlet distribution (\ref{eq:Dir_distr}) with hyperparameters $\alpha_i=1$ using the library NumPy in Python. We believe that our method should work regardless of the chosen prior distribution. For the aforementioned figures, the physical error rates $\langle \lambda_i\rangle_{P_\text{phy}}$ are obtained by considering a small perturbation of the prior expected values. Specifically, each $\langle \lambda_i\rangle_{P_\text{phy}}=\langle \lambda_i\rangle_{P^{(0)}}+\chi$ can be modified by adding a random number $\chi$ sampled from the uniform distribution with support in $[-\delta,\delta]$ and by subsequently renormalizing the expected values such that their sum is $\sum_i \langle \lambda_i\rangle_{P_\text{phy}}=1$. In our case we consider $\delta=0.02$, additionally, all the figures utilize the same physical and prior error rates.
\newline
\indent To account for the errors introduced by the noisy flag gadgets (see Section~\ref{sec:noisyGadgets}), we utilize a circuit-level error model of strength $p$, were each component of a quantum circuit is followed or preceded by a noisy process, see Fig.~\ref{fig:errors_general}. Concretely, for the preparation of noisy $|+\rangle$ states, we model them as noiseless $|+\rangle$ states followed by a phase flip ($Z$) error occurring with probability $p$. Similarly, for noisy $X$-type measurements, we model them as noiseless $X$-type measurements preceded by a phase flip error occurring with probability $p$. Finally, for two qubit entangling gates between data and ancilla qubits, we model them as noiseless gates followed by a depolarizing error channel with strength $p$,
\begin{equation}
\quad \quad\mathcal{D}_2(\rho)=(1-p)\rho +\sum_{P\in\ \mathcal{P}_2/\{I\otimes I\}}\frac{p}{15}\:P\rho P\:,
\end{equation}
where $\mathcal{P}_2=\{I,X,Y,Z\}^{\otimes 2}$ is the Pauli group acting on $2$-qubits.
\newline
\indent The updated expected values $\langle \lambda_i\rangle_{P^{(n)}}$ are computed using equation (\ref{eq:expected_4gadgets}) for Fig. \ref{fig:4layers_CNOT}, \ref{fig:4layers_CNOT_noisy} and~\ref{fig:noisy_4layers_vsp}, and equation (\ref{eq:approx_second_ord}) for Fig.~\ref{fig:approx_second_order}. All the simulations utilize $\alpha_0=2000$. The number of sampled shots is: $n=10000$ for Fig.~\ref{fig:4layers_CNOT} and \ref{fig:4layers_CNOT_noisy}, $n=1000$ for Fig.~\ref{fig:approx_second_order} and $n\in [1,20000]$ for Fig.~\ref{fig:noisy_4layers_vsp}. The error bars represent one standard deviation obtained from the square root of the variance from equation (\ref{eq:var_4layered}). Finally, on Fig.~\ref{fig:noisy_4layers_vsp} the TVD distance is computed using equation (\ref{eq:distance}).


\end{document}